\documentclass[journal]{IEEEtran}
\ifCLASSINFOpdf
\else
\fi
\usepackage[normalem]{ulem}
\usepackage{amsmath,amssymb,amsfonts}
\usepackage{algorithmic}
\usepackage{textcomp}
\usepackage{xcolor}
\usepackage{theorem}
\usepackage{multirow}
\usepackage[acronym]{glossaries}
\usepackage{makecell}
\usepackage{url}
\usepackage{graphicx}
\usepackage{subfigure}

\begin{document}
%
\title{When the Metaverse Meets Carbon Neutrality:\\Ongoing Efforts and Directions}
%
%
%

\author{Fangming~Liu*,~\IEEEmembership{Senior~Member,~IEEE},
        Qiangyu~Pei,
        Shutong~Chen,
        Yongjie~Yuan,
        Lin~Wang,~\IEEEmembership{Senior~Member,~IEEE},
        and Max M{\"u}hlh{\"a}user,~\IEEEmembership{Senior~Member,~IEEE}
\IEEEcompsocitemizethanks{
\IEEEcompsocthanksitem This work was supported in part by the NSFC under Grant 61761136014 and in part by the National Program for Support of Top-notch Young Professionals in National Program for Special Support of Eminent Professionals. Lin Wang was supported in part by DFG Collaborative Research Center 1053 MAKI B2. (Corresponding author: Fangming Liu)
\IEEEcompsocthanksitem F. Liu is with Huazhong University of Science and Technology, and Peng Cheng Laboratory, China. E-mail: fangminghk@gmail.com
\IEEEcompsocthanksitem Q. Pei, S. Chen, and Y. Yuan are with the National Engineering Research Center for Big Data Technology and System, the Services Computing Technology and System Lab, Cluster and Grid Computing Lab in the School of Computer Science and Technology, Huazhong University of Science and Technology, 1037 Luoyu Road, Wuhan 430074, China. E-mail: peiqiangyu@hust.edu.cn, shutongcs@gmail.com, jayayuan@outlook.com
\IEEEcompsocthanksitem L. Wang is with VU Amsterdam, The Netherlands and TU Darmstadt, Germany. E-mail: lin.wang@vu.nl
\IEEEcompsocthanksitem M. M{\"u}hlh{\"a}user is with the Telecooperation Lab, TU Darmstadt, Germany. E-mail: max@informatik.tu-darmstadt.de}}

%
%

\markboth{Journal of \LaTeX\ Class Files,~Vol.~14, No.~8, December~2022}%
{Shell \MakeLowercase{\textit{et al.}}: Bare Demo of IEEEtran.cls for IEEE Journals}
%



\maketitle

\begin{abstract}
The metaverse has recently gained increasing attention from the public. It builds up a virtual world where we can live as a new role regardless of the role we play in the physical world. However, building and operating this virtual world will generate an extraordinary amount of carbon emissions for computing, communicating, displaying, and so on. This inevitably hinders the realization of carbon neutrality as a priority of our society, adding heavy burden to our earth. In this survey, we first present a green viewpoint of the metaverse by investigating the carbon issues in its three core layers, namely the infrastructure layer, the interaction layer, and the economy layer, and estimate their carbon footprints in the near future. Next, we analyze a range of current and emerging applicable green techniques for the purpose of reducing energy usage and carbon emissions of the metaverse, and discuss their limitations in supporting metaverse workloads. Then, in view of these limitations, we discuss important implications and bring forth several insights and future directions to make each metaverse layer greener. After that, we investigate green solutions from the governance perspective, including both public policies in the physical world and regulation of users in the virtual world, and propose an indicator Carbon Utility (CU) to quantify the service quality brought by an user activity per unit of carbon emissions. Finally, we identify an issue for the metaverse as a whole and summarize three directions: (1) a comprehensive consideration of necessary performance metrics, (2) a comprehensive consideration of involved layers and multiple internal components, and (3) a new assessing, recording, and regulating mechanism on carbon footprints of user activities. Our proposed quantitative indicator CU would be helpful in regulating user activities in the metaverse world.
\end{abstract}


%
\IEEEpeerreviewmaketitle

\section{Introduction}\label{sec:intro}

The metaverse concept has attracted considerable attention recently. The term originates from the 1992 science fiction novel \textit{Snow Crash} by Neal Stephenson~\cite{metaverse}; it is a portmaneau of ``meta'' and ``universe'', and denotes a universal and immersive virtual world~\cite{metaverse-wiki}. Creating such a virtual world involves many technologies from artificial intelligence (AI) to extended reality (XR), and many tech companies like Nvidia and Microsoft have paid increasing attention to the metaverse based on their competitive technologies from computing hardware to application platforms. Recently, the COVID-19 pandemic increased the demand for both professional and social online interaction. A prominent example is that the number of daily active users of Roblox, a representative metaverse platform, has grown by 85\% in 2020~\cite{peters-2021-roblox}. 

\begin{figure}[t]
	\centering
	\includegraphics[width=1\linewidth]{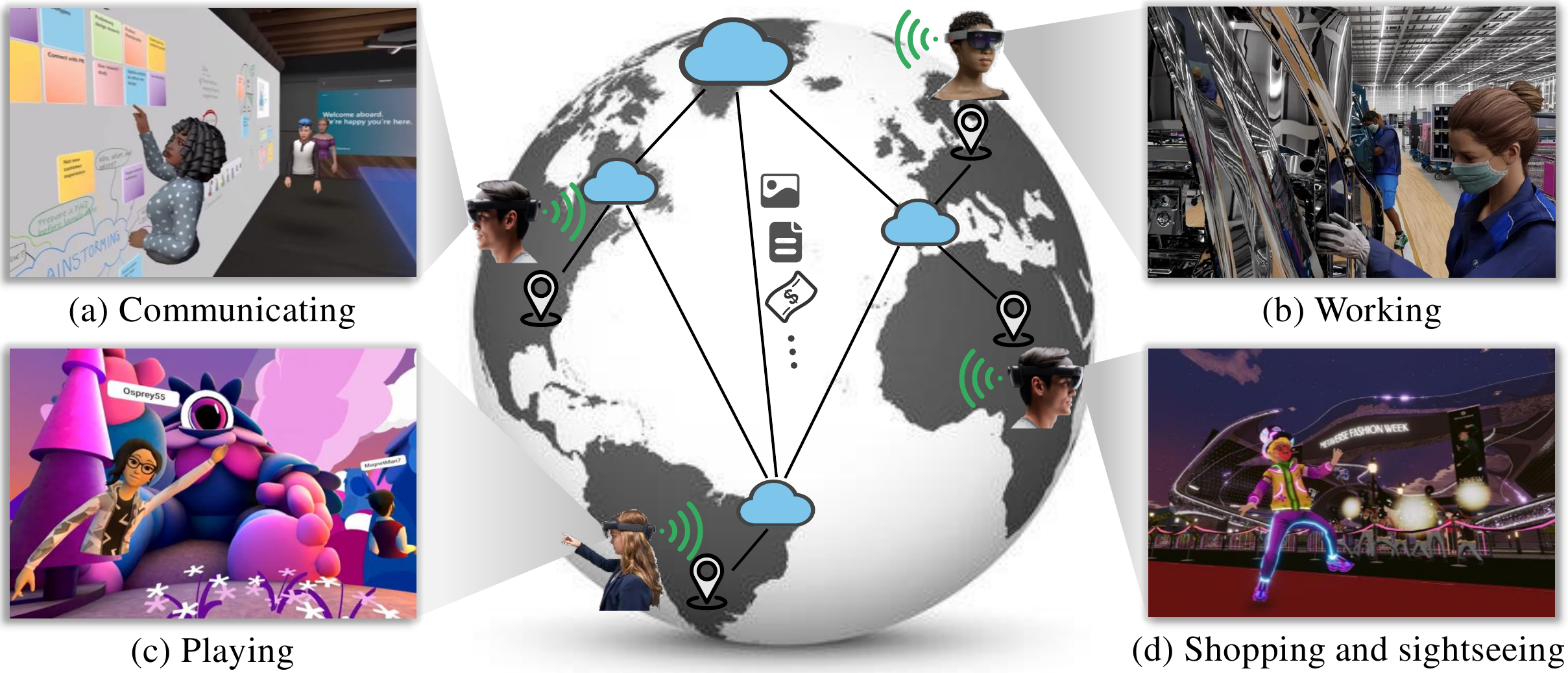}
	\caption{Users from all over the physical world are connected together into the metaverse to communicate, work, play, and more~\cite{ImmersiveView, NVIDIABMW, HorizonWorlds, FashionWeek, GlobeModel, MicrosoftHoloLens, Person1}.}
	\label{fig:MetaverseIllustration}
\end{figure}

The metaverse platforms build virtual worlds for users to connect with others directly online but in an immersive manner over the physical world. Such platforms support extensive daily activities like communication, work, entertainment, trading, etc., as shown in Figure~\ref{fig:MetaverseIllustration}. 
Specifically, Microsoft has developed a virtual collaboration platform Microsoft Mesh where everyone can interact with each other using their avatars while avoiding travelling across the physical world. Users can wear special end-devices running mixed reality (MR) applications, such as HoloLens~2~\cite{MicrosoftHoloLens}, to achieve lifelike experience. 
Recently, the BMW Group and Nvidia together built a highly complex manufacturing system with the Nvidia Omniverse platform~\cite{NvidiaOmniverse, BMW1}. In this virtual factory, engineers can simulate and operate the factory collaboratively in real time before putting the factory into practice in the physical world, which contributes to 30\% more efficient planning processes~\cite{BMW2}. 
Apart from working online, users can play together in this virtual world. In October 2021, Roblox held its first metaverse music festival \textit{World Party}, bringing immersive music experiences to users from all over the world~\cite{RobloxWorldParty}. 
In addition, we can even trade virtual real estate in this virtual world. For example, an anonymous user bought a land next to the famous rapper Snoop Dogg in the metaverse platform The Sandbox with his non-fungible tokens~\cite{SnoopDoggNeighbour}. 
Fortune predicts that the virtual real estate has the potential to reach several trillion dollars in the metaverse, showing the huge demands for trading~\cite{DigitalLand}. In summary, with the rapid development of the metaverse, we can really enjoy a life online as we wish.

\begin{table}[t]
\begin{minipage}[!t]{0.45\columnwidth}
	\renewcommand{\arraystretch}{1.5}
	\caption{Year to reach carbon neutrality of selected countries~\cite{CarbonNeutrality}.}
    \label{tab:CarbonNeutrality}
    \centering
    \begin{tabular}{|c|c|}
        \hline
        \textbf{Country}
        & \textbf{Year}  \\ \hline
        Finland & 2035  \\ \hline
        Iceland & 2040  \\ \hline
        Germany & 2045  \\ \hline
        the US  & 2050  \\ \hline
        the UK  & 2050  \\ \hline
        China   & 2060  \\ \hline
        India   & 2070  \\ \hline
    \end{tabular}
\end{minipage}
\hfill
\begin{minipage}[!t]{0.45\columnwidth}
	\renewcommand{\arraystretch}{1.5}
    \caption{Year to reach carbon neutrality of selected big techs~\cite{CarbonNeutrality2, CarbonNeutrality3, AlibabaCarbonNeutrality}.}
    \label{tab:CarbonNeutrality2}
    \centering
    \begin{tabular}{|c|c|}
        \hline
        \textbf{Company}
        & \textbf{Year}  \\ \hline
        Google  & 2007  \\ \hline
        Netflix & 2022  \\ \hline
        Apple   & 2030  \\ \hline
        Microsoft & 2030  \\ \hline
        Alibaba & 2030  \\ \hline
        Amazon  & 2040  \\ \hline
        Nestl\'e  & 2050  \\ \hline
    \end{tabular}
\end{minipage}
\end{table}

\begin{figure*}[t]
	\centering
	\includegraphics[width=1\textwidth]{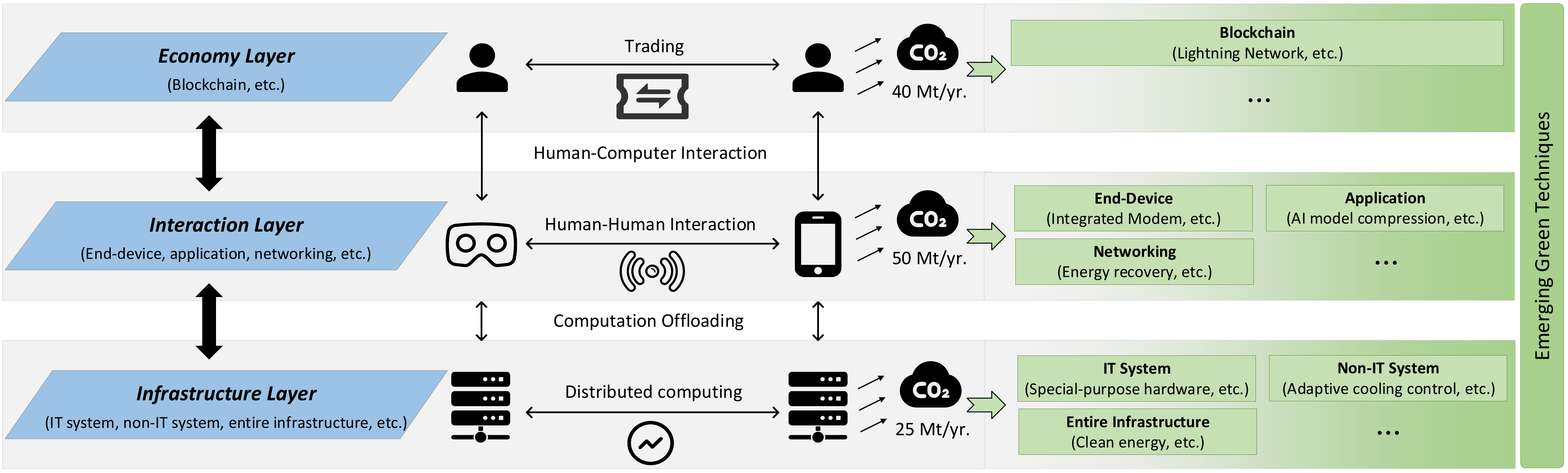}
	\caption{The framework of the metaverse, and the existing green techniques for reducing carbon emissions.}
	\label{fig:Metaverse}
\end{figure*}

It is generally believed that living online is low-carbon since we can avoid activities like flying across the world and holding physical meetings. These physical activities can generate large amounts of energy consumption and produce lots of carbon emissions. For example, an hour flying comes with around 250 kg CO$_{2}$ equivalent based on an estimation~\cite{AviationCarbon}, and the global carbon emissions produced by aviation in 2020 were reduced by 45\% compared to that in 2019 as a result of the COVID-19 pandemic~\cite{tollefson2021covid}.
In addition, according to a recent study, shifting physical meetings to virtual ones can reduce the carbon footprint by as high as 94\%~\cite{tao2021trend}. Nevertheless, although the metaverse provides the potential of decarbonization through limiting physical activities, the carbon emissions will still ``burst'' for supporting the metaverse itself. For example, video streaming is a key technique to the metaverse which provides end-users with a high quality of experience with augmented reality (AR), virtual reality (VR), and MR technologies. However, an estimation shows that four hours of streaming high-quality video (e.g., High Definition (HD) or Ultra HD) a day would result in 53 kg CO$_{2}$ equivalent monthly~\cite{obringer2021overlooked}. The carbon issue can get worse when it comes to transactions. To support transactions in the metaverse, the blockchain technology is believed to be indispensable in the metaverse. However, it is astonishing that an Ethereum transaction can generate 329,000$\times$ more carbon emissions than a credit card transaction currently~\cite{EthereumvsVisa}. In summary, with the development of the metaverse, considerably increasing carbon emissions would be produced in the foreseeable future.

%




Facing the serious consequences of global warming and climate change, countries and enterprises have successively put forward the target of carbon neutrality that refers to achieving net-zero carbon dioxide emissions. Tables~\ref{tab:CarbonNeutrality}~and~\ref{tab:CarbonNeutrality2} list the carbon neutrality goals of selected countries and big techs. As we can see, most developed countries have promised to achieve carbon neutrality by the year of 2050, while many tech companies have promised the same no later than 2030. Those techs have already taken various effective actions to become greener. As early as 2016, Alipay’s Ant Forest initiative was launched intentionally to make everyone take part in low-carbon daily activities such as walking and bicycling instead of driving, by rewarding them with virtual ``green energy points'' that can be translated into real trees thereafter~\cite{AntForest}. Google has been contributing to achieving carbon-free energy supply anytime and anywhere in the world and co-launched a 24/7 Carbon-Free Energy Compact~\cite{CarbonMovement} in 2021. Although there have been a range of actions to achieve carbon neutrality, the rapid development of the metaverse will bring a critical issue on the way to carbon neutrality as a result of the 
continuously growing demands for computing power. To realize a green metaverse, therefore, we raise the following questions:

\begin{itemize}
    \item How much energy will be consumed and how many carbon emissions will be produced by the metaverse?~(\S\ref{sec:framework})
    \item What are the major components of the metaverse and what are their energy consumption and carbon footprint shares? (\S\ref{sec:framework})
    \item Which green techniques can help drive decarbonization of these components, and how is their applicability when dealing with metaverse workloads? (\S\ref{sec:infra}, \S\ref{sec:interaction}, and \S\ref{sec:payment})
    \item What are the limitations of these existing techniques? (\S\ref{sec:infraChallenge}, \S\ref{sec:interactionChallenge}, and \S\ref{sec:paymentChallenge})
    \item What can we do to realize a green metaverse for the target of carbon neutrality? (\S\ref{sec:infraChallenge}, \S\ref{sec:interactionChallenge}, \S\ref{sec:paymentChallenge}, \S\ref{sec:economic} and \S\ref{sec:conclusion})
\end{itemize}
To answer these questions, we examine the carbon issue of the metaverse, provide a comprehensive study of green techniques for the metaverse, provide several insights, and propose future directions tailored to the metaverse. 

The rest of this survey is organized as follows. Section~\ref{sec:framework} provides an overview of the metaverse, comprehensively investigates the carbon issue of the metaverse from the perspective of its three layers, and briefly discuss ongoing efforts of green techniques. The detailed analysis of those green techniques and their remaining challenges are presented in Section~\ref{sec:infra}, Section~\ref{sec:interaction}, and Section~\ref{sec:payment}.  Section~\ref{sec:economic} provides a governance perspective for a low-carbon metaverse. We conclude this survey and present three research directions to a greener metaverse in Section~\ref{sec:conclusion}.


\section{A Green Viewpoint of the Metaverse Framework}\label{sec:framework}

In this section, we first provide an overview of the metaverse and define its three carbon-intensive layers. Then, we analyze the carbon issue of building and operating the metaverse by estimating the carbon emissions of those layers in the following years. Finally, we briefly discuss the ongoing efforts to achieve a green metaverse.

\subsection{An Overview of the Metaverse and Its Three Carbon-Intensive Layers}\label{sec:framework:carbon}

The metaverse refers to a digital virtual world where users gather together to work, study, play, trade, etc., beyond the physical world where we really live. It typically involves two stages, i.e., an offline stage to build the virtual world, and an online stage to operate the virtual world where users can conduct their daily activities. To build the virtual world, large computations are necessarily performed, including AI training, AI inference-based tasks like image recognition and style transfer, 3D rendering, and so on, in order to capture and record key elements from the physical world, convert them into video frames, and render all those frames to construct a 3D virtual world. As an offline process, there is no strong demand for processing latency. After the construction process, users can use their end-devices with avatars to connect to the virtual world and interact with each other. Those end-devices collect all the action information of users from on-board sensors, process them locally or offload them to cloud/edge platforms when the computations become heavy, and generate and render new video frames to display on the end-devices again. As an online process, the end-to-end latency is critical when determining where to perform these computations including AI inference, 3D rendering, video analytics, economic transactions, and so on. 
There is no doubt that performing the aforementioned computations relies on many technologies. A recent study on the metaverse summarizes eight key technologies to support the metaverse, namely hardware infrastructure, network, cloud/edge, AI/blockchain, computer vision, Internet of things/robotics, user interactivity, and extended reality~\cite{lee2021all}. 

From the perspective of functionality and carbon sources of the metaverse, we further propose a metaverse framework with three relatively independent layers as shown in Figure~\ref{fig:Metaverse}, namely the \emph{infrastructure} layer, the \emph{interaction} layer, and the \emph{economy} layer, each of which is composed of several components. Here, we briefly introduce their functionality and main components as follows, and present their carbon footprints in Section~\ref{sec:framework:value}.

(1)~The infrastructure layer is key to supporting \emph{computation} in the metaverse in the form of datacenters. As end-devices like mobile phones and headsets typically have limited computing capacity, huge amounts of computations need to be offloaded to datacenters with powerful computing servers and abundant resources. The datacenter facility typically consists of information technology (IT) and non-IT equipment. Specifically, the IT equipment provides the computing, networking, and storage capability, while the non-IT facility is to support the operation of datacenters, such as the cooling system.

(2)~The interaction layer is key to supporting \emph{communication} between users in the metaverse and includes human-computer and human-human interaction. On the one hand, the human-computer interaction involves both hardware and software components. The former consist of various end-devices like mobile phones and headsets, and the latter refer to diverse kinds of applications running on top of end-devices, such as virtual conferencing and virtual manufacturing. 
On the other hand, the human-human interaction involves the networking equipment, including cellular modem chips on end-devices and cellular base stations, which provides essential supports for multiplayer interaction in the metaverse. 

(3)~The economy layer is key to supporting \emph{transactions} between users in the metaverse. To provide enough transaction security as in the physical world, the economy layer has widely incorporated the blockchain technology. On top of the blockchain, cryptocurrency is a necessary medium for secure transactions between users, while non-fungible tokens are used to prove the ownership rights of users' virtual properties. 

Note that such a division covers all the necessary technologies for the metaverse as introduced at the beginning: (1)~hardware infrastructure, edge/cloud computing, AI, and computer vision covered in the infrastructure layer, (2)~hardware infrastructure, network, AI, computer vision, Internet of things/robotics, user interactivity, and extended reality covered in the interaction layer, and (3)~AI and blockchain covered in the economy layer. Some of the technologies like AI belong to more than one layer because of their comprehensive functionality.

\subsection{The ``Carbon'' Role of the Metaverse}\label{sec:framework:value}

According to Emergen Research~\cite{MetaverseMarket}, the global metaverse market size has reached \$63 billion in 2021 and is estimated to reach \$1607 billion in 2030. With the rapid evolvement of the metaverse, there will inevitably be high carbon emissions at the same time. In order to undertake effective measures on carbon reduction, we must firstly understand the ``carbon'' role of the metaverse by speculating its carbon emissions in the following years. 

However, it is nontrivial to conduct the estimation accurately as the metaverse is still a newly emerging and constantly evolving concept that covers a wide range of technologies. To reduce the estimation error as much as possible, we speculate the energy consumption and carbon emissions of the metaverse based on the energy figures provided by the following reputable sources: (1) the global IT market size and metaverse market size: a report from The Business Research Company~\cite{ITMarket} and a report from Emergen Research~\cite{MetaverseMarket}, respectively, (2) the energy consumption of datacenters, end-devices, and networking: articles on Information and Communications Technology (ICT) energy from Huawei Technologies~\cite{andrae2020new, andrae2015global}, and (3) the energy consumption of the blockchain: a research from the University of Cambridge for calculating the Bitcoin energy~\cite{CryptocurrencyCambridge} and an article from the Technical University of Munich for speculating the overall energy of cryptocurrencies~\cite{gallersdorfer2020energy}. 
Note that we take the energy consumption of cryptocurrencies as that of the blockchain since the blockchain is basically used for transactions in the metaverse. Since the energy numbers of cryptocurrencies are only accessible before 2022, we use an exponential function to estimate the future energy consumption based on historical values~\cite{zade2019bitcoin}. 

\begin{figure}[t]
	\centering
	\includegraphics[width=0.8\linewidth]{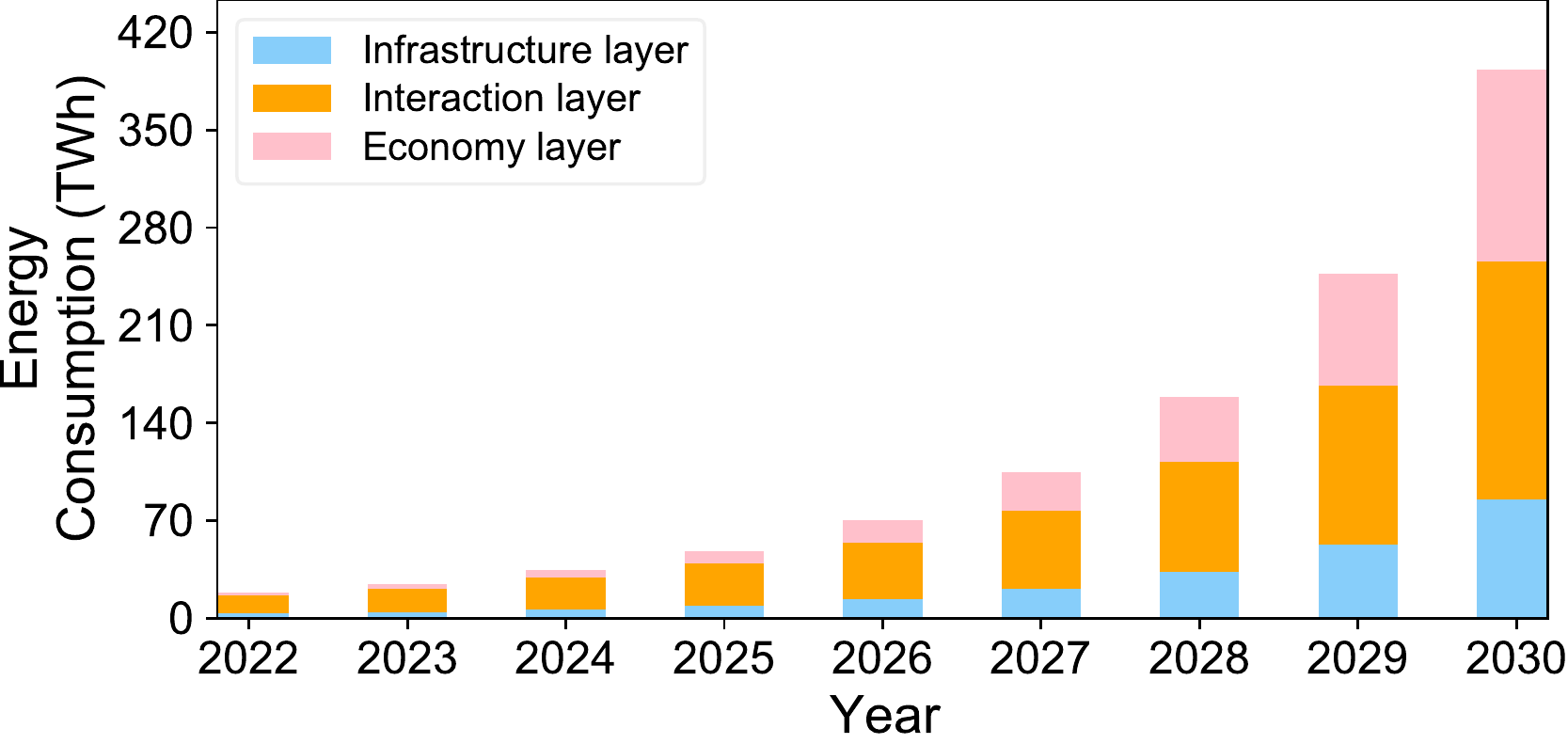}
	\caption{The estimated energy consumption of the metaverse from the year of 2022 to 2030.}
	\label{fig:MetaverseEnergy}
\end{figure}

By multiplying the energy numbers and the ratio of the metaverse market size to the global IT market size, we can estimate the global energy consumption of the metaverse in each of the three layers from the year of 2022 to 2030. Figure~\ref{fig:MetaverseEnergy} plots the results and growth trend. Based on the estimation, we find that the energy consumption of the infrastructure layer grows relatively proportional to the total energy of the metaverse, and occupies about one fifth all the time. By contrast, the energy consumed by the interaction layer remains at a high level constantly. Although a single end-device usually consumes only several Watts, their huge quantity and a growing demand for data transmission still lead to a high energy share for supporting immersive interaction. Meanwhile, we can see that its energy share, however, drops from about three quarters in 2022 to less than half in 2030. The dropped energy share can be attributed to the faster growth of the economy layer, whose energy consumption increases by over 8$\times$ during the eight years. This strong growth is what we have expected, as the current blockchain technology is generally believed to be extraordinarily energy-consuming for the need to perform lots of redundant and useless hash computations. 

\begin{figure}[t]
	\centering
	\includegraphics[width=0.8\linewidth]{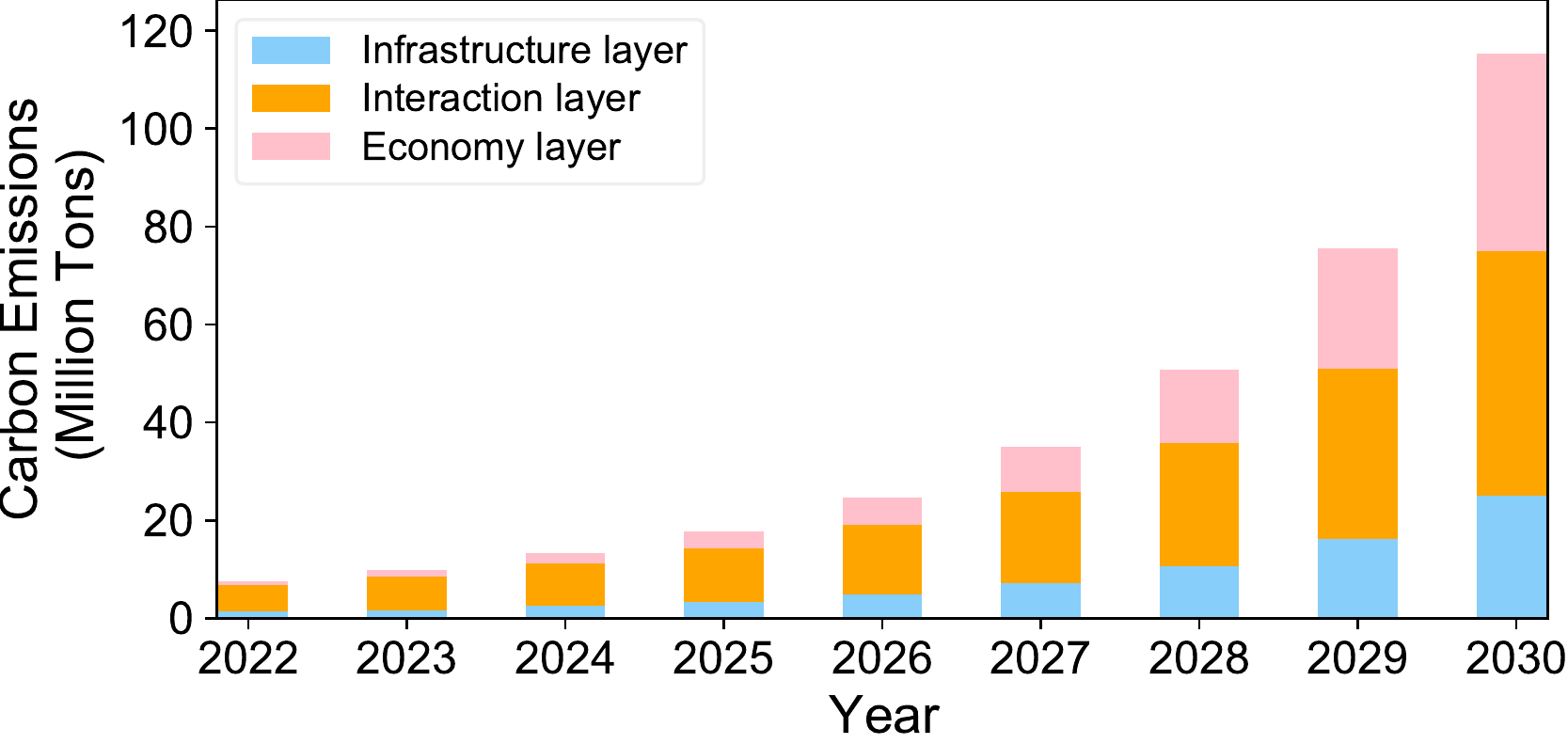}
	\caption{The estimated carbon emissions of the metaverse from the year of 2022 to 2030.}
	\label{fig:MetaverseCarbon}
\end{figure}

To quantify the carbon emissions, we should keep in mind that they are not only determined by the energy consumption, but also positively correlated with the carbon intensity of electricity generation which gets decreased due in large part to an increasing share of clean energy. 
Therefore, based on the energy numbers shown in Figure~\ref{fig:MetaverseEnergy} and the carbon intensity of electricity generation~\cite{CarbonIntensity}, we estimate the corresponding carbon emissions of each layer from 2022 to 2030 as depicted in Figure~\ref{fig:MetaverseCarbon}. We find that the carbon emissions of the metaverse will reach as high as 115.30~Mt by the year of 2030. To combat the climate change and achieve carbon neutrality as promised, global CO$_{2}$ emissions should be reduced to 23.63 Gt by 2030~\cite{freitag2021real}, in which circumstance the metaverse accounts for nearly 0.5\% of the global carbon emissions! 
We argue that it is increasingly urgent to consider enough green aspects when building and operating the metaverse and devise various green techniques to tackle its carbon issue.

\textbf{\underline{Implications}}~\textit{The metaverse will account for as high as 0.5\% of the global carbon emissions by 2030 unless we take effective preventive measures from now on. First of all, we should call for global attention to its carbon issue and encourage collaborative efforts on green techniques.}




\subsection{An Overview of Ongoing Green Efforts for the Metaverse}

To address the carbon issue in the IT sector, many green techniques have emerged. As presented in Figure~\ref{fig:Metaverse}, these techniques will help reduce the carbon footprint in the three layers of the metaverse as well. At the infrastructure layer, the techniques can be categorized into the improvement of the IT system, cooling system, power system, as well as the entire infrastructure concerning all of them. At the interaction layer, given the whole process of interaction with others, the techniques include the improvement of end-devices, AR/VR/MR-based applications, and networking. At the economy layer, the techniques mainly refer to the improvement of the blockchain technology. Nevertheless, these existing techniques also show limitations when dealing with metaverse workloads. In the following sections, we will dive into these green techniques for the three metaverse layers by analyzing their applicability and limitations. It is worth noting that our estimation to the energy consumption and carbon emissions of the metaverse in Section~\ref{sec:framework:value} would not influence the discussion on green techniques in the following sections. Even if there exists a relatively high estimation error, in view of the growing trend of the metaverse, the carbon footprint of the metaverse will still be large enough to emphasize the importance of green efforts. 


\section{How can the Infrastructure Become Green?}\label{sec:infra}

\begin{table*}[t]
	\renewcommand{\arraystretch}{1.5}
    \centering
    \caption{Green Techniques for the Infrastructure.}
    {
    \begin{tabular}{|cc|l|l|}
    \hline
    \multicolumn{2}{|c|}{\textbf{Level}}                                         & \multicolumn{1}{c|}{\textbf{Green technique}} &  \multicolumn{1}{c|}{\textbf{Work}} \\ \hline
    \multicolumn{1}{|c|}{\multirow{7}{*}{IT system}} & \multirow{5}{*}{\makecell*[c]{Component-level \\ (Sec.~\ref{sec:infraITcomponent})}} & Replacing instruction set architectures     & RISC~\cite{RISC}    \\ \cline{3-4} 
    \multicolumn{1}{|c|}{}                           &                            & In-memory processing & TIMELY~\cite{li2020timely}                            \\      \cline{3-4} 
    \multicolumn{1}{|c|}{}                           &                            & Deploying special-purpose hardware & Graviton3~\cite{AmazonGraviton3}, IBM~\cite{agrawal20217nm}                            \\      \cline{3-4} 
    \multicolumn{1}{|c|}{}                           &                            & Dynamic voltage frequency scaling &
    Rubik~\cite{kasture2015rubik}, Gemini~\cite{zhou2020gemini}               \\ \cline{3-4} 
    \multicolumn{1}{|c|}{}                           &                            & AI model compression & SmartExchange~\cite{zhao2020smartexchange}, OFA~\cite{cai2019once}             \\ \cline{2-4}
    \multicolumn{1}{|c|}{}                           & \multirow{2}{*}{\makecell*[c]{Server-level \\ (Sec.~\ref{sec:infraITserver})}}    & Single-server workload scheduling &  FineStream~\cite{zhang2020finestream}, Wiesner et al.~\cite{wiesner2021let}                             \\ \cline{3-4} 
    \multicolumn{1}{|c|}{}                           &                            & Cross-server workload scheduling & Yi et al.~\cite{yi2019toward}, Poly~\cite{wang2019poly}                             \\ \hline
    \multicolumn{1}{|c|}{\multirow{5}{*}{Non-IT system}} & \multirow{3}{*}{\makecell*[c]{Cooling system \\ (Sec.~\ref{sec:infranonIT:cooling})}} & Choosing suitable cooling techniques &  GreenEdge~\cite{zhou2019greenedge}, CoolEdge~\cite{pei2022cooledge}         \\ \cline{3-4} 
    \multicolumn{1}{|c|}{}                           &                            & Adaptive cooling control & CoolEdge~\cite{pei2022cooledge}, DeepEE~\cite{ran2019deepee}           \\ \cline{3-4} 
    \multicolumn{1}{|c|}{}                           &                            & Cooling-aware workload management & DeepEE~\cite{ran2019deepee}, CoolAir~\cite{nguyen2015coolair}               \\ \cline{2-4}
    \multicolumn{1}{|c|}{}                           & \multirow{2}{*}{\makecell*[c]{Power system \\ (Sec.~\ref{sec:infranonIT:power})}}    & Reducing power mismatches & HEB~\cite{liu2015heb}, Flex~\cite{zhang2021flex}                         \\ \cline{3-4} 
    \multicolumn{1}{|c|}{}                           &                            & Reducing power losses   & Panama Power~\cite{PanamaPower}                 \\ \hline
    \multicolumn{2}{|c|}{\multirow{3}{*}{\makecell*[c]{Entire infrastructure (Sec.~\ref{sec:infraEntire})}}}                  & Improving clean energy usage & Gupta et al.~\cite{gupta2019combining}, UFC~\cite{zhou2014fuelcell}                    \\ \cline{3-4} 
    \multicolumn{2}{|c|}{}                                                        & Heat harvesting &  CloudHeat~\cite{chen2018cloudheat}, H2P~\cite{zhu2020heat}                        \\ \cline{3-4} 
    \multicolumn{2}{|c|}{}                                                        & Emerging computing technologies & CryoWire~\cite{min2022cryowire}, Zuchongzhi~\cite{wu2021strong}                      \\ \hline
    \end{tabular}}
    \label{tab:InfraGreenTechniques}
\end{table*}

Datacenters are the key infrastructure for the metaverse by providing powerful computing capacity~\cite{ComputingPower, lee2021all}. As presented in Section~\ref{sec:framework:carbon}, those large computations from AI-based tasks to 3D rendering have a huge appetite for computing power. However, due to the limited computing and battery capacity of end-devices, it is hard for them to provide real-time metaverse services with a high quality-of-experience (QoE). A promising solution is to offload most of the computations to the datacenter infrastructure though incurring network latency. For example, 3D rendering is a critical technique for generating the visual content of the metaverse, which requires large amounts of computing power especially when rendering a high-quality frame (e.g., 4K Ultra HD)~\cite{jiang2021reliable} and thus may need to be conducted in datacenters. In the following, we will first introduce the carbon issue of the infrastructure and then analyze green techniques to improve the energy efficiency of the IT systems, non-IT systems, and the entire infrastructure. In the end, we provide future directions to become green.

\begin{figure}[t]
	\centering
	\includegraphics[width=0.92\columnwidth]{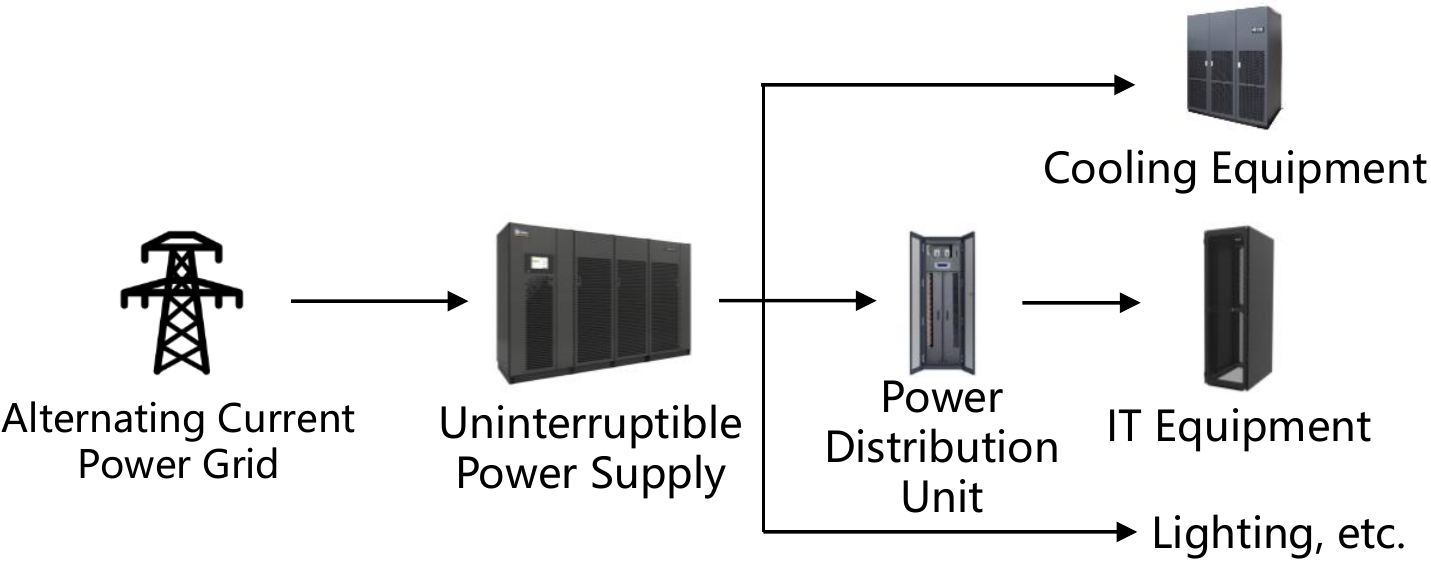}
	\caption{Datacenter equipment.}
	\label{fig:Datacenter}
\end{figure}

\subsection{The Carbon Issue of the Infrastructure}

A datacenter is typically composed of the IT and non-IT equipment~\cite{WhatDataCenter}, as illustrated in Figure~\ref{fig:Datacenter}. The IT equipment includes servers, network switches, etc., for providing the computation, network transmission, and storage capacity; the non-IT equipment is to support the IT equipment, referring to the power system, cooling system, fire suppression system, etc. Currently, the global datacenter electricity consumption has already reaches as high as 200 TWh, around 1\% of the worldwide electricity consumption and 0.3\% of global carbon emissions~\cite{jones2018stop}. With the growing demand of the metaverse and the growing computing power of global datacenters, it is vital to improve the energy efficiency and reduce the carbon footprint of the energy-hungry and carbon-intensive datacenter infrastructure~\cite{chien2021driving}. 

\begin{figure}[t]
	\centering
	\includegraphics[width=0.9\columnwidth]{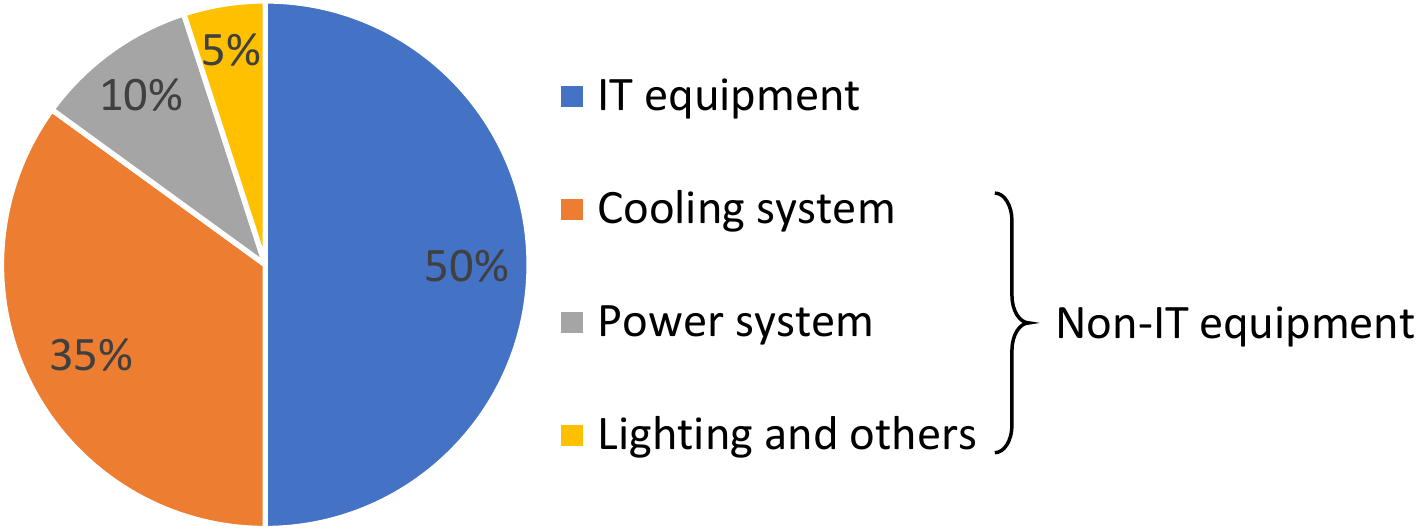}
	\caption{The energy share by source in a typical datacenter.}
	\label{fig:EnergyShare}
\end{figure}

To this end, we first investigate how to measure the energy efficiency of datacenters. 
Figure~\ref{fig:EnergyShare} shows the share of energy consumption by source in a typical datacenter~\cite{EnergyFigure}. We can find that as the most important part of a datacenter, IT equipment consumes about 50\% of the total energy, while among the non-IT equipment, the cooling system plays the dominant role in energy consumption. The power usage effectiveness (PUE) is a commonly used indicator for measuring the energy efficiency of a datacenter as a whole, and is defined as the ratio of the total amount of energy used by the whole datacenter infrastructure to the energy delivered to the IT equipment~\cite{PUE}. It is reported that the energy efficiency of datacenters was generally poor a decade ago: the average PUE in 2011 was as high as 1.89. With the development of modern cooling techniques (e.g., water cooling with warm water~\cite{jiang2019fine}, free cooling under the sea~\cite{MicrosoftDC}) and power supply (e.g., renewable energy), the PUE of large-scale datacenters have dropped significantly in recent years, e.g., 1.06 by Google~\cite{GoogleEfficiency}. 

\begin{figure}[t]
	\centering
	\includegraphics[width=1\columnwidth]{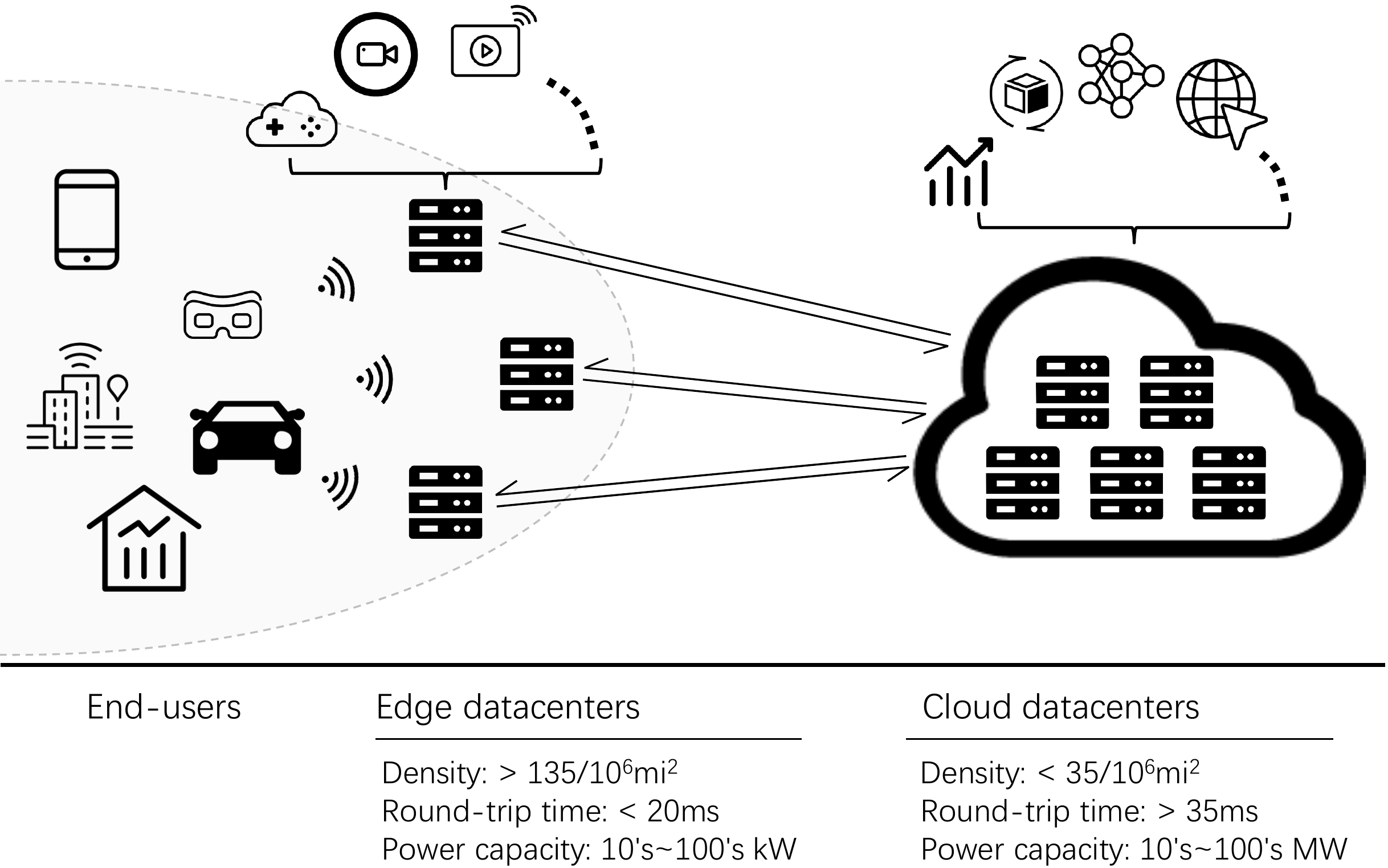}
	\caption{An overview of the datacenter infrastructure for the metaverse.}
	\label{fig:CloudAndEdge}
\end{figure}

According to the distance to end-users and the type of provided services, datacenters can be generally classified into edge datacenters and cloud datacenters, both of which play an important role in the metaverse~\cite{lee2021all}. Specifically, edge datacenters are mainly built for hosting online services while cloud datacenters focus on delay-tolerant offline services. To provide low-latency and high-throughput services to end-users, edge datacenters need to be widely distributed across the network edge near the urban. However, because of the area restriction in the urban, the edge datacenters own limited computing capacity but higher computing density~\cite{pei2022cooledge, VaporChamber}. By contrast, cloud datacenters own massive computing and storage capacity by locating in rural areas, making them cover a wider population but get distant from end-users. A cloud datacenter generally contains thousands to tens of thousands of server racks. Its power rating can reach 10's to 100's of~MW~\cite{RackNumber}, which is three orders of magnitude higher than that of an edge datacenter. 
Figure~\ref{fig:CloudAndEdge} summarizes the differences in provided services, location, and computing capacity, between edge datacenters and cloud datacenters~\cite{xu2021cloud, StateOfTheEdge-report2021}.

Because of these distinctions, cloud and edge datacenters have different levels of energy efficiency and face different opportunities and challenges in carbon reduction. For example, edge service providers sometimes need to over-provision resources to satisfy the strict QoE requirement~\cite{xu2021cloud}, which in turn leads to extra energy consumption. Therefore, although energy efficient solutions for cloud datacenters have been well studied over the past few decades, some green techniques devised for cloud datacenters may become invalid in edge datacenters considering their distinctions and the carbon issue becomes more serious. 
For instance, since the free cooling technique is almost unavailable, the PUE of edge datacenters is just as high as~2~\cite{EdgePUE}. 

\begin{figure}[t]
	\centering
	\includegraphics[width=1\columnwidth]{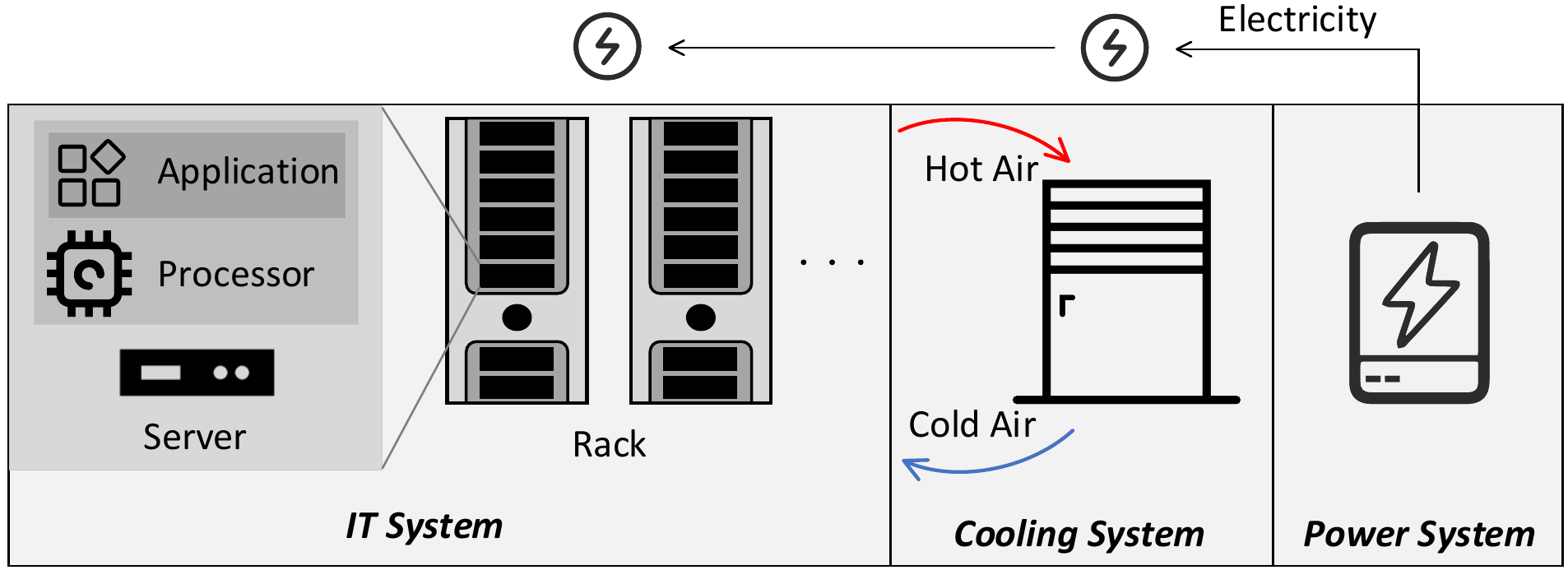}
	\caption{Key components for carbon reduction of the datacenter infrastructure.}
	\label{fig:Infrastructure}
	\vspace{8pt}
\end{figure}

The serious carbon issue pushes us to develop green techniques to reduce carbon emissions from the datacenter infrastructure. As listed in Table~\ref{tab:InfraGreenTechniques}, based on the datacenter architecture and carbon sources, there are extensive green techniques for improving energy efficiency and reducing carbon emissions in multiple levels. Figure~\ref{fig:Infrastructure} shows the key components for carbon reduction of the datacenter infrastructure. 

\subsection{Energy Efficiency Improvement of IT Systems}\label{sec:infraIT}

The energy efficiency of IT systems mainly depends on computing servers and each individual component inside servers. We summarize the energy efficiency improvement at the component level and the server level separately.

\subsubsection{Component-level energy efficiency improvement}\label{sec:infraITcomponent}

To process a certain task efficiently, datacenters have gradually deployed newly-developed special-purpose hardware (e.g., GPU and FPGA) nowadays in addition to CPU~\cite{MetaverseGPU}. Component-level green techniques focus on a single component, including CPU, GPU, memory, and other accelerators through hardware and/or software design. 

\textbf{Replacing instruction set architectures.} The evolving of instruction set architectures (ISAs) is a promising way for energy saving. The reduced instruction set computer (RISC) architecture~\cite{RISC} (e.g., used in ARM processors) is recognized as a more energy-efficient ISA than the conventional complex instruction set computer (CISC) architecture (e.g., used in $\times$86 processors). Because of this advantage, RISC has been widely adopted in end-devices with limited battery capacity. Recently, a growing number of datacenter operators have turned to ARM-based servers for higher efficiency. For example, Cloudflare deployed its edge servers with ARM CPUs (RISC) in 2021 that enables 57\% more Internet requests per Watt than the performance achieved by the latest generation of edge servers equipped with AMD Rome CPUs (CISC)~\cite{ArmCPU}. Leveraging RISC-based processors, the metaverse service provider can handle more high-concurrency requests while keeping a small energy footprint in the coming metaverse era. 

\textbf{In-memory processing.} Apart from being compute-intensive, a lot of metaverse services are also memory-intensive. For example, to enhance the immersive experience, some metaverse applications like cloud gaming will include human-computer interaction using voice, where non-player characters can chat with real players to provide next-step instructions. Therefore, speech recognition is a key technique in the human-computer interaction, but it often incurs a high memory footprint and consumes a large amount of energy~\cite{han2017ese}. For these memory-intensive metaverse services, it is critical to optimize the memory footprint. 
Processing-in-memory (PIM) is an emerging technique to improve energy efficiency by reducing data movement between memories and CPUs. Li et al.~\cite{li2020timely} propose a resistive-random-access-memory (ReRAM) based PIM accelerator with three hardware-based designs to save energy. 

\textbf{Deploying special-purpose hardware.} To serve as a foundation for the computational metaverse, various newly-developed special-purpose hardware products like Nvidia A100 GPU~\cite{A100} are promising to be widely deployed in datacenters in addition to general-purpose CPUs~\cite{MetaverseGPU}. These special-purpose hardware components that cater to the characteristics of a certain type of tasks often achieve a high energy efficiency. For example, Amazon introduces its 3rd generation Graviton CPU (Graviton3) for its cloud datacenters. It is reported that Graviton3 saves up to 60\% energy compared to Graviton2 probably due to the integrated special-purpose processors~\cite{AmazonGraviton3}. With an increasing demand for AI-based services, the industry has already developed a variety of special-purpose hardware that supports low-precision computations of AI workloads to relieve the resource requirement and improve the energy efficiency, such as the NNP I-1000 accelerator designed by Intel~\cite{IntelNNP}. 
IBM has been empowering CPU chips with AI capability by integrating special-purpose AI chips and aims to boosting energy efficiency by 2.5$\times$ each year~\cite{IBMChip}. Recently, it designs an AI chip with the 7nm technology which supports low-precision computations and achieves up to 16.5 TOPS/W as compared to 3.12 TOPS/W achieved by the widely-used Nvidia A100 datacenter GPU~\cite{agrawal20217nm}. 
In addition to hardware designs, most of these components are also integrated with a power management unit, such as FIVR~\cite{burton2014fivr}, to achieve a dynamic and desired trade-off between energy consumption and performance. 
These special-purpose components are especially promising for capacity-limited edge datacenters that provide various special metaverse services with stringent performance constraints. 

\begin{figure}[t]
	\centering
	\includegraphics[width=0.88\columnwidth]{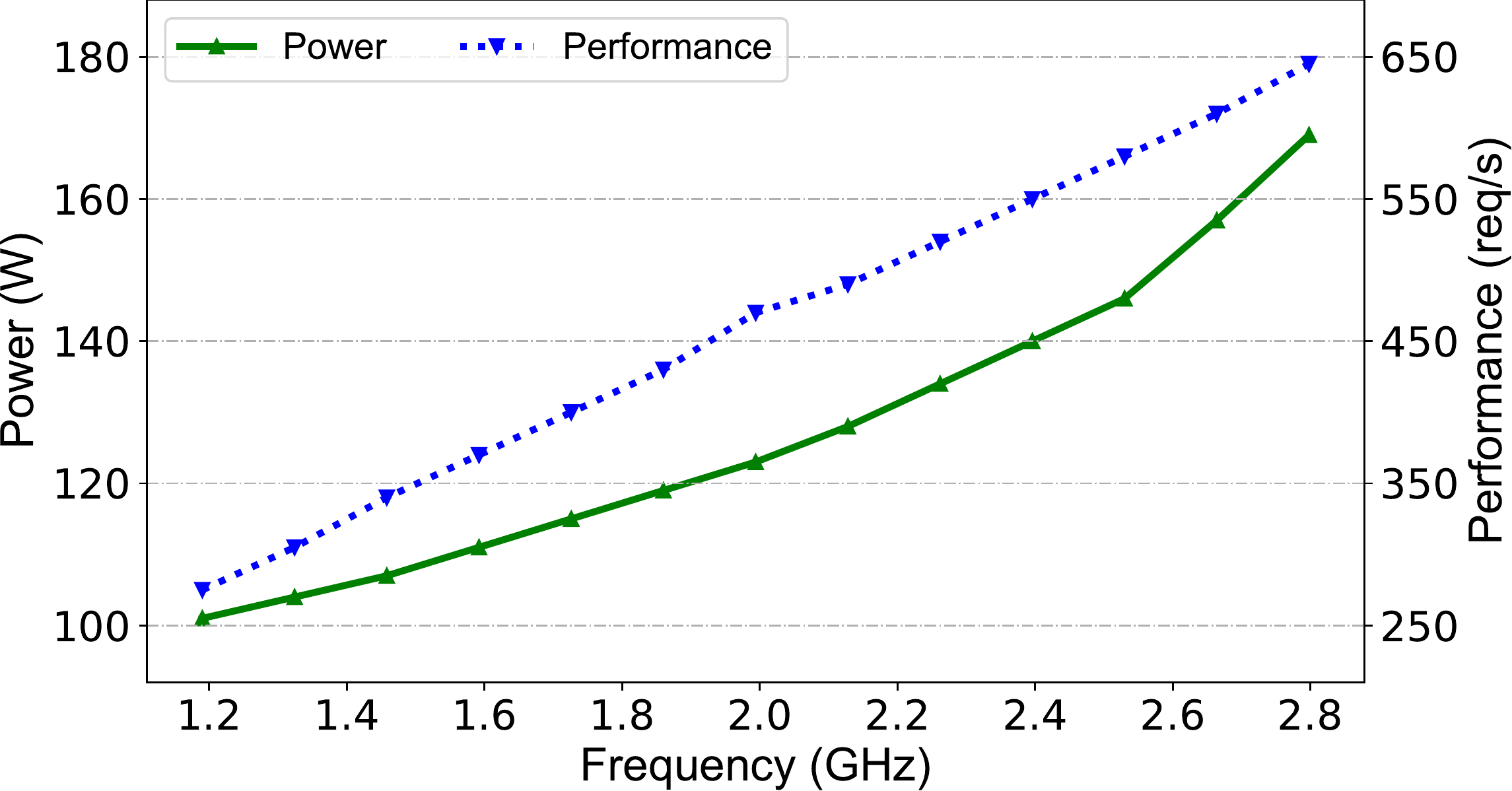}
	\caption{The relationship between the frequency and power, and the frequency and performance of the Intel Core i5 processor.}
	\label{fig:PowerVoltage}
\end{figure}

\textbf{\underline{Implications}}~\textit{It is especially essential to deploy special-purpose hardware for both performance and energy efficiency of the metaverse. However, given the various metaverse workloads, metaverse operators need to carefully choose the best-suitable hardware combinations. Moreover, as such hardware usually brings high capital expenditures, operators should also keep in mind that special-purpose hardware is not a silver bullet compared to more economical CPUs.}

\textbf{Dynamic voltage frequency scaling.} Dynamic voltage frequency scaling (DVFS) is a widely-used software-based techniques to improve the energy efficiency of components. 
Figure~\ref{fig:PowerVoltage} illustrates that the power footprint of the Intel Core i5 Processor is positively correlated with its frequency~\cite{petrucci2011optimized}. Based on this observation, DVFS is advanced to reduce the power consumption by dynamically adjusting the voltage and frequency of components like CPUs~\cite{kasture2015rubik, zhou2020gemini} and other special-purpose hardware including GPUs~\cite{santriaji2016grape}, application specific integrated circuits (ASICs)~\cite{chen2015execution}, field programmable gate arrays (FPGAs)~\cite{chen2015execution}, etc.
However, Figure~\ref{fig:PowerVoltage} also shows that a higher energy efficiency usually comes at the expense of performance. It would be more appropriate to implement DVFS carefully on delay-tolerant services, such as background image re-rendering in the offline stage, rather than time-sensitive services like cloud gaming and online meeting. The metaverse service providers should accurately estimate the service execution time and carefully balance the energy efficiency and QoS violation rate~\cite{wang2013impact}. 

\textbf{\underline{Implications}}~\textit{DVFS is a widely-used technique to save energy. However, metaverse operators need to carefully incorporate this technique as it usually trades performance for energy efficiency, while metaverse applications tend to place strict requirements. Particularly, even slight performance violations may cause a serious consistency issue as a result of wrong time order of consecutive events in the virtual world.}

\textbf{AI model compression.} It is generally believed that the AI-based services play a key role in the metaverse. Nevertheless, their huge energy consumption deserves our special attention as well. For example, when training a large AI model containing 6 billion parameters only to 13\% of the whole process, the carbon emission emitted by GPUs will be as much as powering a home for one year in America~\cite{trainingcarbonemission}. The model compression technique has emerged as an effective approach to reduce carbon emissions of model training and inference, by decreasing the size of models and thus the resource and time overhead without significantly affecting the accuracy. 

Common model compression techniques include quantization, sparsification, tensor decomposition, and so on~\cite{deng2020model}. 
Model compression can help reduce energy consumption of AI-based services by not only relaxing the demand for computing resources, but also eliminating large amounts of data transferring and storage. 
An effective model compression result usually comes with the hardware software co-design that jointly considers the accelerator architecture. 
Xia et al.~\cite{xia2016switched} propose a quantization method to reduce the amount of intermediate data and eliminate the power-consuming Digital-to-Analog and Analog-to-Digital Converters on newly designed special-purpose hardware, namely resistive random-access memory (ReRAM). 
Zhao et al.~\cite{zhao2020smartexchange} propose an algorithm-hardware co-design framework for energy-efficient inference. Using pruning, decomposition, and quantization jointly, the proposed framework reduces the energy consumption in data movement and weight storage. 

The model compression technique is also leveraged when distributing a trained model to various end-devices. To provide AI-based services for end-devices with heterogeneous architectures, conventional approaches usually train a specialized model for each, which causes prohibitive computation and energy overhead. Cai et al.~\cite{cai2019once} propose a once-for-all network and utilize the pruning method to obtain a sub-network from the once-for-all network for each end-device, which reduces carbon emissions by up to 1300$\times$ through avoiding additional training.


\subsubsection{Server-level energy efficiency improvement}\label{sec:infraITserver}

In the server level, energy-aware or carbon-aware workload scheduling is key to increasing the energy efficiency. Workload scheduling is to decide ``when, where, and how'' the tasks scheduled in server(s) can achieve the least energy consumption and carbon emissions. 
There are generally two ways to schedule workloads: within a single server and across servers. 

\textbf{Single-server workload scheduling.} The service provider can apply resource allocation, workload consolidation, priority-based scheduling, and so on, to manage the server's energy consumption and corresponding carbon emissions. The single-server workload scheduling generally improves the energy efficiency at the expense of the performance. For example, decreasing the quota of computation resources (e.g., GPU memory) for the object detection task will save more energy while hurting the accuracy, which would degrade QoE of the AR service greatly~\cite{liu2018dare}. The metaverse service provider should carefully balance the trade-off between performance and energy efficiency, especially for real-time metaverse services. 
Prekas et al.~\cite{prekas2015energy} propose a latency-critical workload consolidation scheme combined with the DVFS technique for improving energy efficiency. Zhang et al.~\cite{zhang2020finestream} focus on the emerging CPU-GPU integrated architecture that eliminates frequent communications between CPU and GPU through low-bandwidth and high-latency PCI-e. They advance a fine-grained data stream system to schedule workload between CPU and GPU, and achieve 1.8$\times$ energy reduction as compared to the stream system based on the original discrete architecture. Wiesner et al.~\cite{wiesner2021let} propose a carbon-aware workload scheduling that prioritizes time-sensitive workloads and postpones delay-tolerant ones until the energy source is less carbon-intensive, such as the abundant solar energy during daylight hours.

\textbf{Cross-server workload scheduling.} The service provider would need to place and migrate workloads across servers considering the difference among servers in resource utilization, resource interference, energy efficiency, etc. 
The performance of the metaverse services often behaviors differently on heterogeneous hardware. For example, the inference time of YOLOv3, a commonly-used object detection model, deployed on Nvidia Jetson Nano is 20$\times$ longer than that on Nvidia GTX1060~\cite{lee2021yolo}. The metaverse service provider should capture these heterogeneous behaviors and provide sufficient performance guarantees during the cross-server workload scheduling. 
Yi et al.~\cite{yi2019toward} design a workload allocation algorithm for long-lasting and compute-intensive tasks. Based on the deep reinforcement learning technique, the proposed algorithm captures the complex power and thermal dynamics of servers, and learns to make efficient workload allocation decisions that achieve a high power efficiency. 
Wang et al.~\cite{wang2019poly} propose a QoE-aware workload scheduling framework to reduce energy consumption. The proposed framework jointly utilizes GPU- and FPGA-based accelerators to cap peak server loads during a day, which reduces the energy consumption by up to 23\% with QoE guarantees. 
Considering the trend that the ISA architecture is evolving in datacenters, Barbalace et al.~\cite{barbalace2017breaking} develop an efficient workload scheduling solution to migrate executions between heterogeneous-ISA servers. The proposed solution can reduce over half of the original energy consumption.

\textbf{\underline{Implications}}~\textit{Workload scheduling is a common technique used in datacenters. To schedule various metaverse workloads, service providers need to carefully characterize and categorize these workloads in advance to reach an efficient balance between performance and energy consumption at run time.}




\subsection{Energy Efficiency Improvement of non-IT Systems}\label{sec:infranonIT}

Besides the energy consumption of IT systems, attention should be paid on the energy efficiency of cooling systems and power systems that are the most power-consuming parts among non-IT systems.

\subsubsection{Cooling systems}\label{sec:infranonIT:cooling}
In datacenters, power flows through the IT systems and is transformed into thermal energy, which should be discharged by cooling systems to ensure the safety of IT systems. High-demand metaverse services not only consume a large amount of IT energy, but also increase the pressure of cooling systems.

There are three commonly-used cooling techniques in datacenters: air cooling, liquid cooling, and free cooling. Air cooling is the most widely-used cooling technique in datacenters since it can be deployed in whatever environment. However, air cooling is also the most energy-consuming one because of the low heat conduction capacity of the air, difficulty in controlling the air flow, and so on. Designing a thermal-aware workload scheduling approach is an intuitive way to reduce the energy consumption and improve the efficiency of air cooling in the era of the metaverse. 
However, it is hard to capture the system dynamics and complexity since metaverse service requests cannot be easily predicted and the cooling energy is affected by multiple factors such as the workload placement and airflow rates. To handle the system dynamics and complexity, Ran et al.~\cite{ran2019deepee} propose a deep reinforcement learning based solution to reduce the cooling energy through workload scheduling and cooling adjustment, which achieves up to 15\% of the energy savings. 

Compared with air cooling, liquid cooling has better efficiency because of the higher density and thermal capacity of the liquid like water. 
Liquid cooling can be classified as direct-to-chip cooling and immersion cooling. In a direct-to-chip cooling system, cold liquid removes heat from components through a cold plate which is directly pressed on the surface of each component. While in a immersion cooling system (either single-phase or two-phase), all components are submerged in a thermally conductive dielectric liquid to dissipate the heat. Jalili et al.~\cite{jalili2021cost} conduct a comprehensive study on two-phase immersion cooling in terms of performance, costs, etc., which lowers the datacenter PUE by about 14\% than that of an air-cooled datacenter. Zhou et al.~\cite{zhou2019greenedge} discuss the energy inefficiency in edge datacenters, and suggest some future trends, such as prioritizing the cooling technique of immersion liquid cooling. However, compared to the direct-to-chip cooling, immersion cooling places strict constraints on the fluid, tank, etc., which increases the capital expenditure and may prevent large-scale adoption especially in edge datacenters~\cite{pei2022cooledge}. 
According to a comprehensive analysis of edge datacenters, Pei et al.~\cite{pei2022cooledge} find that existing cooling techniques are inefficient for edge datacenters because of the proximity, power density, and diverse edge services. To address the challenges of heterogeneity and high density of edge datacenters, they advance a fine-grained and custom-designed warm water cooling solution with high energy efficiency. 

It is worth noting that many natural cooling sources can be leveraged to cool datacenters without additional energy consumption, such as cold air and water. Free cooling aims to make use of external low ambient temperature to dissipate the heat without the refrigeration process~\cite{FreeCooling}. To this end, free-cooled datacenters should be sited in cold locations for access to the free cooing sources. However, there are also some challenges to utilize free cooling. The temperature fluctuation of natural cooling sources during the day would probably reduce the hardware reliability~\cite{manousakis2016environmental}. Besides, the effect of workload placement on the heat recirculation should be also carefully considered to reduce the temperature variation. To handle the above challenges, Nguyen et al.~\cite{nguyen2015coolair} propose a joint management of workloads and cooling control for free-cooled datacenters to maintain the air temperature with low cooling energy. 

\textbf{\underline{Implications}}~\textit{The cooling efficiency is highly dependent on specific datacenter infrastructures and workloads. Thus, we should differentiate between different datacenters that focus on different metaverse services and accordingly adopt a targeted and even customized cooling technique.}

\begin{table*}[t]
	\renewcommand{\arraystretch}{1.5}
	\caption{The Carbon Intensity of Major Energy Sources (In Descending Order in Each Category)~\cite{CarbonIntensitybySource}}
    \label{tab:CleanEnergy}
    \centering
    \begin{tabular}{|c|ccc|c|cccccl|}
    \hline
    \multirow{2}{*}{\textbf{Energy source}}         & \multicolumn{3}{c|}{\makecell*{Conventional non-renewable \\ fossil fuels}} & \makecell*{Other non-renewable \\ sources} & \multicolumn{6}{c|}{Renewable sources}                                                                                                               \\ \cline{2-11} 
                                                    & \multicolumn{1}{c|}{Coal}                     & \multicolumn{1}{c|}{Oil}                   & Natural Gas                   & Nuclear                                                                                        & \multicolumn{1}{c|}{Solar} & \multicolumn{1}{c|}{Geothermal} & \multicolumn{1}{c|}{Biomass} & \multicolumn{1}{c|}{Wind} & \multicolumn{2}{c|}{Hydro} \\ \hline
    \textbf{Carbon intensity (g~CO$_{2}$~eq/kWh)} & \multicolumn{1}{c|}{1,001}                    & \multicolumn{1}{c|}{840}                   & 469                   & 16                                                                                             & \multicolumn{1}{c|}{22-48} & \multicolumn{1}{c|}{45}         & \multicolumn{1}{c|}{18}      & \multicolumn{1}{c|}{12}   & \multicolumn{2}{c|}{4}     \\ \hline
    \end{tabular}
\end{table*}


\begin{figure*}[t]
	\centering
	\includegraphics[width=0.75\linewidth]{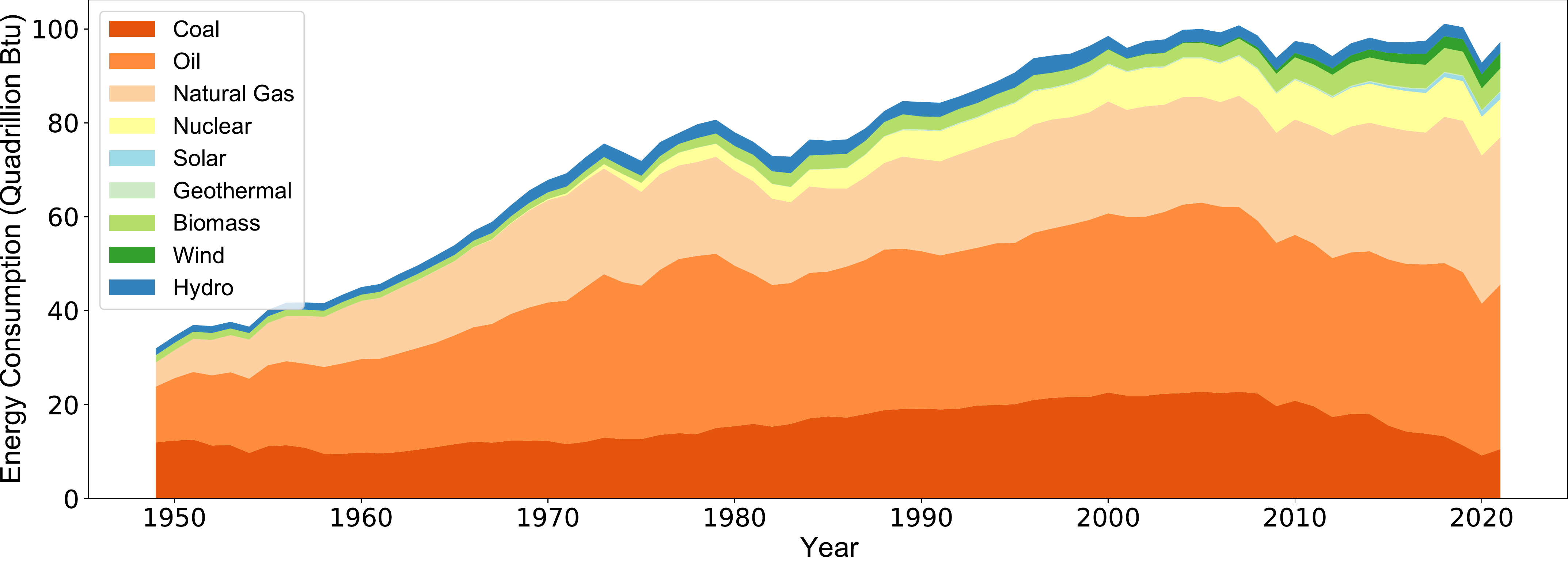}
	\caption{Non-Renewable and Renewable Energy Consumption by Source in the United States from 1949 to 2021~\cite{EnergybySource}.}
	\label{fig:EnergyConsumptionSource}
\end{figure*}

\subsubsection{Power systems}\label{sec:infranonIT:power}
Power supply and distribution systems are essential to daily reliable operations of datacenters. It is of great importance to ensure high availability of the power systems for providing 7$\times$24 metaverse services. To this end, many datacenter operators tend to equip their datacenters with redundant power and cooling capacity. Zhang et al.~\cite{zhang2021flex} point out that about 10\%-50\% of the power resources are typically reserved, leading to a waste of energy and extra expenses. To improve the energy efficiency while ensuring 7$\times$24 reliable services, it is necessary to reduce power mismatches with hardware- or software-based solutions. The hardware-based solutions are to utilize non-IT equipment, such as energy storage, to balance the power demand and ensure the availability. Liu et al.~\cite{liu2015heb} propose to combine super-capacitors with traditional uninterruptible power supply systems, which helps improve clean energy usage and leads to a high energy efficiency. Shen et al.~\cite{shen2013achieving} focus on the power management of hardware components, i.e., microprocessors and peripherals, which achieves minimum power consumption under performance constraints. The software-based solutions are to achieve peak shaving through workload management. Zhang et al.~\cite{zhang2021flex} introduce ``zero-reserved-power'' datacenters with availability guarantees. Specifically, they find that there are some software-redundant workloads (i.e., Software-as-a-Service-based workloads like Web search concerning that they are always replicated across multiple availability zones~\cite{zhang2021flex}) that can tolerate infrastructure failures, such as 3D rendering in the offline stage. Utilizing these workloads, they propose a workload management solution to reduce the reserved resources and hence decrease the redundant power consumption while ensuring the system availability. 

Power losses of the power systems during energy conversion is another important problem in datacenters as it consumes around 10\% of total datacenter energy. There are several energy-efficient solutions on various components of power systems, such as uninterruptible power supplies (UPS), transformers, and power distribution units~\cite{PowerLoss}. Recently, Delta and Alibaba introduce a new power supply system with the Panama Power solution that achieves 2\% $\sim$ 4.6\% energy efficiency improvement by transforming 10kV alternating current to 240V/336V direct current directly~\cite{PanamaPower}.

\subsection{Energy Efficiency Improvement of the Entire Infrastructure}\label{sec:infraEntire}

\subsubsection{Clean energy}\label{sec:infraEntireClean}

The adoption of clean energy can greatly reduce the carbon footprint from the aspect of energy supply because of its much lower carbon intensity than conventional fossil energy sources, as listed in Table~\ref{tab:CleanEnergy}.  Figure~\ref{fig:EnergyConsumptionSource} plots their energy shares by source in the United States from 1949 to 2021~\cite{EnergybySource}. We can find a constantly growing share of renewable energy sources, especially after the year of 2007 when the total fossil energy consumption peaks and starts to decline. To combat global warming, it is reported that the clean energy (e.g., renewable energy, nuclear energy, and fossil energy with carbon capture and storage technologies) would need to contribute 70\% of the energy demand in 2050~\cite{CleanEnergyReport}. As a pioneer of green computing, Google achieved carbon neutral as early as 2007, and aims to be carbon free in 2030. The continuous promotion of clean energy plays a key role where Google matches the global energy consumption with 100\% renewable energy since 2017~\cite{GoogleCarbon}. 

There also exist many researches demonstrating that the clean energy has great potential for increasing the energy efficiency of datacenters. Deng et al.~\cite{deng2014harnessing} comprehensively study why, when, where, and how to use renewable energy in datacenters. They highlight the necessity to match multiple energy source supply with variable power demands. Gupta et al.~\cite{gupta2019combining} investigate the potential to combine multiple clean energy sources for complementary benefits. 
Zhou et al.~\cite{zhou2014fuelcell} analyze the benefits of fuel cell generation in geo-distributed datacenters for the first time. As the output of the fuel cell is tunable, it can be adjusted to match the time-varying energy demand of datacenters and avoid energy waste. 

However, it is challenging for datacenters to use clean energy when managing metaverse services. 
The generation of renewable electricity highly depends on the weather, and thus the clean energy supply is uncertain, intermittent, and variable~\cite{deng2014harnessing,zhou2016bilateral}. As the metaverse requires 7$\times$24 services and the user requests could be highly dynamic during the online stage, the use of clean energy may cause significant supply-demand mismatches and hurt the reliability of the metaverse. To address these challenges, there are three ways to achieve the service reliability while utilizing the clean energy: predicting the clean power generation, leveraging the energy storage, and utilizing the temporal and/or spatial correlation of the clean energy. 

In the first approach, the service provider can schedule the workload based on the estimation of renewable energy generation. For example, Aksanli et al.~\cite{aksanli2011utilizing} design short-term prediction algorithms for both solar energy and wind energy, and propose a workload scheduling approach to improve the proportion of clean energy usage. However, the prediction accuracy of renewable energy is just passable because of the frequent weather changes~\cite{deng2014harnessing}. Another potential approach is to use the energy storage technology to bridge the supply-demand mismatches. 
Ren et al.~\cite{ren2013carbon} present an algorithm to carefully decide the energy storage's charging/discharging rate and renewable power capacity. 
Deng et al.~\cite{deng2013online} design an online algorithm to decide the UPS's charging/discharging rate and the power procurement considering the time-varying clean energy price, certain volumes of intermittent renewable energy, and the capacity of UPS. The third approach is to utilize the temporal and/or spatial correlation of the clean energy. Some companies build datacenters in a distributed manner for providing a wider range of services. For these geo-distributed datacenters, Zhou et al.~\cite{zhou2013carbon} find that there is a spatial and temporal variability of the electricity's carbon footprint due to the different fuel mixes of electricity generation among different regions and periods. Leveraging the variability, they design a carbon-aware control framework that jointly reduces power costs and carbon emissions through geographical load balancing, capacity right-sizing, and server speed scaling. 

\textbf{\underline{Implications}}~\textit{Clean energy is believed to be promising in reducing carbon emissions. However, since different metaverse applications will match different datacenter types as discussed before, e.g., cloud and edge, the supply-demand imbalance across them can reduce the usage effectiveness of clean energy, where energy transmission and storage techniques may help.}

\subsubsection{Heat harvesting}\label{sec:infraEntireHeat}
As discussed in Sec.~\ref{sec:infranonIT:cooling}, the growing demand of the metaverse not only makes the carbon emissions burst but also increases the pressure of cooling systems. What is more, after the IT systems get cooled, the thermal energy tends to be dissipated to the environment directly, which may sometimes have a negative impact on the environment and is energy-consuming~\cite{chen2018cloudheat}. To fully utilize the heat, waste heat harvesting has been emerging in the practice and literature, which is promising to reduce carbon emissions by enabling other industries to leverage this recycled heat, such as warming buildings in some middle- and high-latitude cities.

There are two common ways to realize district heating depending on the servers' distance to buildings, i.e., locally or remotely. Liu et al.~\cite{liu2011data} propose a concept of ``data furnace'' which delivers heat to residential buildings by placing servers just in them. In recent years, ``data furnace'' has been commercialized, e.g., a computing heater by Qarnot~\cite{Qarnot}. If the servers are far from buildings in the form of datacenters, it is necessary to set up district heating systems to transfer the heat. Chen et al.~\cite{chen2018cloudheat} analyze the feasibility and profitability to harvest heat from datacenters and warm buildings with the help of district heating systems. They propose a market mechanism to motivate datacenter operators to sell waste heat to district heating systems, which not only improves the energy efficiency of datacenters but also reduces the usage of fossil energy in heat generation. To reduce the heat transferring overhead, it is rather practical to recycle heat from edge servers while providing real-time metaverse services, as edge datacenters are typically located in or next to residential/commercial areas. Thanks to the location, the waste-heat recovery system in the Tencent Tianjin datacenter directly brings heating to more than 5,100 local households, saving the high costs for long-haul delivery of the heat~\cite{TencentHeat}. The reduced 52,400 tonnes of carbon dioxide emissions every year is equivalent to the carbon footprint of Portland~\cite{TencentHeat, CarbonFootprint}.

Besides harvesting heat for warming, some researches focus on reusing waste heat to generate electricity. Lee et al.~\cite{lee2015hot,lee2016thermoelectric} leverage thermoelectric generators (TEGs) to recycle heat from microprocessors and power the fans or TECs for cooling. Zhu et al.~\cite{zhu2020heat} focus on warm-water-cooled datacenters, and use TEGs to recycle waste heat from the heated cooling water and convert it into electricity. The power reusing efficiency of the proposed approach reaches 14.23\% on average.


\subsubsection{Emerging computing technologies} There are some newly emerging advanced computing technologies that may contribute to energy efficiency improvement and carbon reduction of the datacenter infrastructure.

\textbf{Cryogenic computing.} Recent researches find that cryogenic computing can reduce the component power by about one order of magnitude without performance degradation~\cite{byun2020cryocore}. However, there could be higher difficulty in managing the cooling system and higher cooling energy consumption to support the extremely low temperature, which may make this technology still be in the pilot phase. Min et al.~\cite{min2022cryowire} further propose a CPU microarchitecture under cryogenic computing. Extensive evaluations indicate that it achieves higher performance while reducing both hardware and cooling energy consumption.

\textbf{Quantum computing.} It is recognized that classic computers consume significant amounts of energy especially for high performance computing. With the evolving of the computing technology, quantum computing is believed to provide higher computing capacity with less energy consumption. By leveraging the superposition property as well as other technologies, such as quantum tunneling, quantum bits are exponentially efficient than classical bits. The recent quantum computer \textit{Zuchongzhi} is estimated to be tens of thousands times faster than the world's most powerful supercomputer by then~\cite{wu2021strong}. Meanwhile, according to an estimation, quantum can reduce energy consumption by hundreds to thousands times~\cite{QuantumComputingEnergy}. In short, quantum computers are continually advanced and considered to revolutionize the global datacenter energy~\cite{QuantumComputing}, which also makes sense to the infrastructure for the metaverse.


\subsection{Directions to Become Green}\label{sec:infraChallenge}
Although there have been extensive studies on green techniques, there are still several challenges in the infrastructure layer. Specifically, we propose an insight into the deployment of special-purpose hardware components in datacenters as these components are becoming increasingly popular to handle specific metaverse workload types. 

\textit{\textbf{Insight:}} \textbf{Special-purpose hardware is not a silver bullet for metaverse computing.} There are many off-the-shelf special-purpose hardware components with different characteristics and they typically achieve a higher energy efficiency when executing certain tasks than general-purpose CPUs. For example, data processing units (DPUs) can perform data processing efficiently while AI accelerators are specially designed for AI tasks. Nonetheless, it is not a good idea to execute all metaverse workloads on these special-purpose hardware components. Firstly, special-purpose hardware components may perform worse than CPUs for some memory-intensive metaverse workloads. A recent work indicates that CPUs achieve much lower inference latency than GPUs when executing memory-intensive AI models like long short-term memory~\cite{le2020allox}. Secondly, special-purpose hardware components are generally more expensive than CPUs. For example, an A100 GPU costs over \$10,000, while a CPU typically costs only a few thousand dollars. Last but not least, as a kind of general-purpose hardware components, CPUs can carry out a wide variety of tasks so that they are more capable of shaving peak computation demands as the demand of different types of workloads fluctuate. 

\textit{\textbf{Direction:}} On the one hand, datacenter operators should have a knowledge of the provided metaverse services of the datacenter and the required computations (e.g., AI inference and 3D rendering). To make full use of both general-purpose and special-purpose hardware components, it is necessary for datacenter operators to comprehensively consider the performance, cost, energy efficiency, etc., to choose customized hardware combination, i.e., the number of each hardware type in a specific datacenter. On the other hand, it is also necessary for service providers to have a deep understanding of the various components in terms of the energy usage, and accurately measure and disaggregate the energy consumption of metaverse workloads in their life cycles on each component. Note that given the changing amounts of clean energy production and total power supply, the most desired power demand can also vary dynamically. For example, the electricity prices in many European countries are even set negative sometimes in order to maintain a must-run power capacity~\cite{EuropePrices}, which indicates that higher energy consumption of datacenters or a lower energy efficiency may be desired in this case. In all, the service providers need to comprehensively consider the performance, energy efficiency, etc., and devise a dynamic service deployment mechanism to schedule various metaverse services to a specific hardware component at run time. 

\section{How can the Interaction Become Green?}\label{sec:interaction}

\begin{table*}[t]
	\renewcommand{\arraystretch}{1.5}
    \centering
    \caption{Green Techniques for the Interaction.}
    \begin{tabular}{|cc|l|l|}
    \hline
    \multicolumn{2}{|c|}{\textbf{Level}}                                         & \multicolumn{1}{c|}{\textbf{Green technique}}                                & \multicolumn{1}{c|}{\textbf{Work}}   \\ \hline
    \multicolumn{1}{|c|}{\multirow{9}{*}{\makecell*[c]{End-device}}} & \multirow{3}{*}{\makecell*[c]{Processor \\ (Sec.~\ref{sec:hardwaredeviceprocessors})}}   & Heterogeneous processing architectures   & ARM big.LITTLE~\cite{ArmCore}            \\ \cline{3-4} 
    \multicolumn{1}{|c|}{}                             &                              & Die shrink        &  2~nm processor~\cite{FirstChip}             \\ \cline{3-4} 
    \multicolumn{1}{|c|}{}                             &                              & Deploying special-purpose processors    & Holographic processor~\cite{MicrosoftHoloLens}    \\ \cline{2-4} 
    \multicolumn{1}{|c|}{}                             & \multirow{4}{*}{\makecell*[c]{Modem chip \\ (Sec.~\ref{sec:networkDevices})}}  & Using integrated modems & Dimensity series~\cite{5GMorePower}                 \\ \cline{3-4} 
    \multicolumn{1}{|c|}{}                             &                              & Dynamic voltage frequency scaling & UltraSave~\cite{5GModems}             \\ \cline{3-4} 
    \multicolumn{1}{|c|}{}                             &                              & Power management & Dynamic search space adaptation~\cite{5GDeviceEfficiency}                 \\ \cline{3-4} 
    \multicolumn{1}{|c|}{}                             &                              & Switching between wireless technologies & Xu et al.~\cite{xu2020understanding}, Narayanan et al.~\cite{narayanan2021variegated}    \\ \cline{2-4} 
    \multicolumn{1}{|c|}{}                             & \makecell*[c]{Others \\ (Sec.~\ref{sec:hardwaredeviceother})}             & Heat harvesting    & Dai et al.~\cite{dai2018exploiting}, Yap et al.~\cite{yap2016thermoelectric}             \\ \hline
    \multicolumn{1}{|c|}{\multirow{8}{*}{Application}} & \multirow{6}{*}{\makecell*[c]{Processing technology \\ (Sec.~\ref{sec:application:processing} and \\ Sec.~\ref{sec:application:video})}} & AI model compression             & Quantization~\cite{Quantization}, pruning~\cite{yang2017designing}      \\ \cline{3-4} 
    \multicolumn{1}{|c|}{}                          &                                          & Collaborative execution & Neurosurgeon~\cite{Neurosurgeon}, JointDNN~\cite{eshratifar2019jointdnn}               \\ \cline{3-4} 
    \multicolumn{1}{|c|}{}                          &                                          & Dynamic voltage frequency scaling & Tang et al.~\cite{tang2019impact}, Nabavinejad et al.~\cite{nabavinejad2019coordinated}                      \\ \cline{3-4} 
    \multicolumn{1}{|c|}{}                          &                                          & Workload scheduling & AsyMo~\cite{wang2021asymo}                   \\ \cline{3-4} 
    \multicolumn{1}{|c|}{}                          &                                          & Customized AI inference frameworks & TensorFlow Lite~\cite{TensorFlow}              \\ \cline{3-4} 
    \multicolumn{1}{|c|}{}                          &                                          & Video-oriented techniques & Palleon~\cite{feng2021palleon}, VR-DANN~\cite{song2020vr}                \\ \cline{2-4} 
    \multicolumn{1}{|c|}{}                          & \multirow{2}{*}{\makecell*[c]{Display technology \\ (Sec.~\ref{sec:application:VRAR})}}    & Reducing unnecessary data transferring & BurstLink~\cite{haj2021burstlink}       \\ \cline{3-4} 
    \multicolumn{1}{|c|}{}                          &                                          & Reducing redundant calculations & Déjà View~\cite{zhao2020deja}, MARLIN~\cite{apicharttrisorn2019frugal}     \\ \hline
   \multicolumn{1}{|c|}{\multirow{4}{*}{Networking}} & \multirow{2}{*}{\makecell*[c]{Base station \\ (Sec.~\ref{sec:networkStations})}} & Architectural design & I et al.~\cite{chih2020energy}    \\ \cline{3-4} 
    \multicolumn{1}{|c|}{}                        &                                                                                                                     & Power management & 5G Power~\cite{Huawei5GPower}     \\ \cline{2-4} 
    \multicolumn{1}{|c|}{}                        & \makecell*[c]{Data transmission \\ (Sec.~\ref{sec:networkStations})}             & Energy recovery & Wu et al.~\cite{wu2017overview}     \\ \hline
    \end{tabular}
    \label{tab:InteractionGreenTechniques}
\end{table*}

In addition to the datacenter infrastructure, both \textit{human-computer interaction} and \textit{human-human interaction} play a key role in presenting immersive experience to users. The human-computer interaction relies on various end-devices and AR/VR/MR-based applications. The end-devices involve mobile phones, glasses, headsets, sensors, and even motion-capture gloves for lifelike experience~\cite{OfficeMeetings}, while the applications include virtual conferencing, cloud gaming, etc., running on top of these end-devices. Instead of a single user, the human-human interaction provides essential supports for the communication of multi-players in the metaverse. Such support relies on both networking components (e.g., cellular modem chips on end-devices and cellular base stations) and data transmission between these components. Considering the requirements of the metaverse for ultra low latency, high bandwidth, high speed, low jitter, etc., the emerging 5G technology is promising to support large-scale human-human interaction by nature as compared with 4G/LTE~\cite{lee2021all}. By means of 5G, it is possible to offload computational metaverse workloads like AI-based tasks from battery-constrained end-devices to datacenters and download other players' data at the same time as discussed in Section~\ref{sec:infra}~\cite{Nokia5G}. In the following, we will first introduce the carbon issue of the interaction and then analyze green techniques to improve the energy efficiency of the end-devices, applications, and networking components. In the end, we provide future directions to become green.

\subsection{The Carbon Issue of the Interaction}\label{sec:InteractionCarbon}
With the evolvement of the metaverse, users are demanding more advanced interaction technologies for ultimate immersive experience. Meanwhile, there will be inevitably huge global energy consumption and carbon emissions of the interaction. Firstly, although the power consumption of a single end-device is only several Watts on average~\cite{XRPower}, the total energy consumption will still be high since the number of end-devices is continually growing, especially for the smartphones and wearable devices like VR headsets. In particular, the number of global mobile devices rises from less than 8 billion in 2016 to more than 12 billion by 2022~\cite{MobileNumber}, while VR and AR headsets are estimated to grow about 10 times from about 11 million in 2021 to 105 million in 2025~\cite{XRNumber}. 
Secondly, the increasing demand for metaverse applications also lead to high energy consumption and carbon emissions. For example, the carbon emissions from streaming workloads in 2018 is almost equivalent to those from France~\cite{StreamingCarbon}. The real-time interaction among users in the metaverse relies heavily on the streaming-based applications and will raise the carbon emissions further. The second example is the popular computer vision and natural language processing tasks, where the energy consumption of one AI inference execution is gradually approaching and has even surpassed the energy consumption of a human body per second, respectively~\cite{desislavov2021compute}. This indicates the severe energy burden by AI inference in view of its widespread use as the metaverse continues expanding to benefit everyone. In summary, it is estimated that the global carbon emissions of end-devices will reach about 442 Mt in 2022, comparable to the carbon emissions of Canada~\cite{andrae2020new, CarbonIntensity, CarbonFootprintbyCountry}. 

Meanwhile, the proliferation of a sheer number of end-devices and applications to support the metaverse will put increasing pressure on the networking energy usage. According to an estimation by Cisco~\cite{MobileNumber}, the global mobile data traffic will increase from 12 exabytes per month in 2017 to 77 exabytes by 2022. Although the energy efficiency of 5G is higher than previous 4G/LTE, the larger antenna numbers, higher base station density, and higher bandwidth~\cite{chih2020energy}, as well as the high data transfer demand by the metaverse will make its total energy consumption increase sharply. According to a recent report~\cite{5GBaseStations}, as compared with a 4G base station, a typical 5G base station increases the energy consumption by up to twice or more. It is estimated that the global carbon emissions of networking will reach about 95 Mt in 2022, comparable to the carbon emissions of Netherlands, and is expected to increase by 2.7$\times$ by 2030~\cite{andrae2020new, CarbonIntensity, CarbonFootprintbyCountry}. On the other hand, when it comes to end-devices supporting higher band frequencies, Redmi claims that 5G mobile phones tend to consume about 20\% more power than 4G ones~\cite{Redmi5G}, and real measurement studies conducted by Xu et al.~\cite{xu2020understanding} and Narayanan et al.~\cite{narayanan2021variegated} also demonstrate the much higher energy consumption of mobile devices over 5G than 4G. 

\begin{figure}[t]
	\centering
	\includegraphics[width=1\columnwidth]{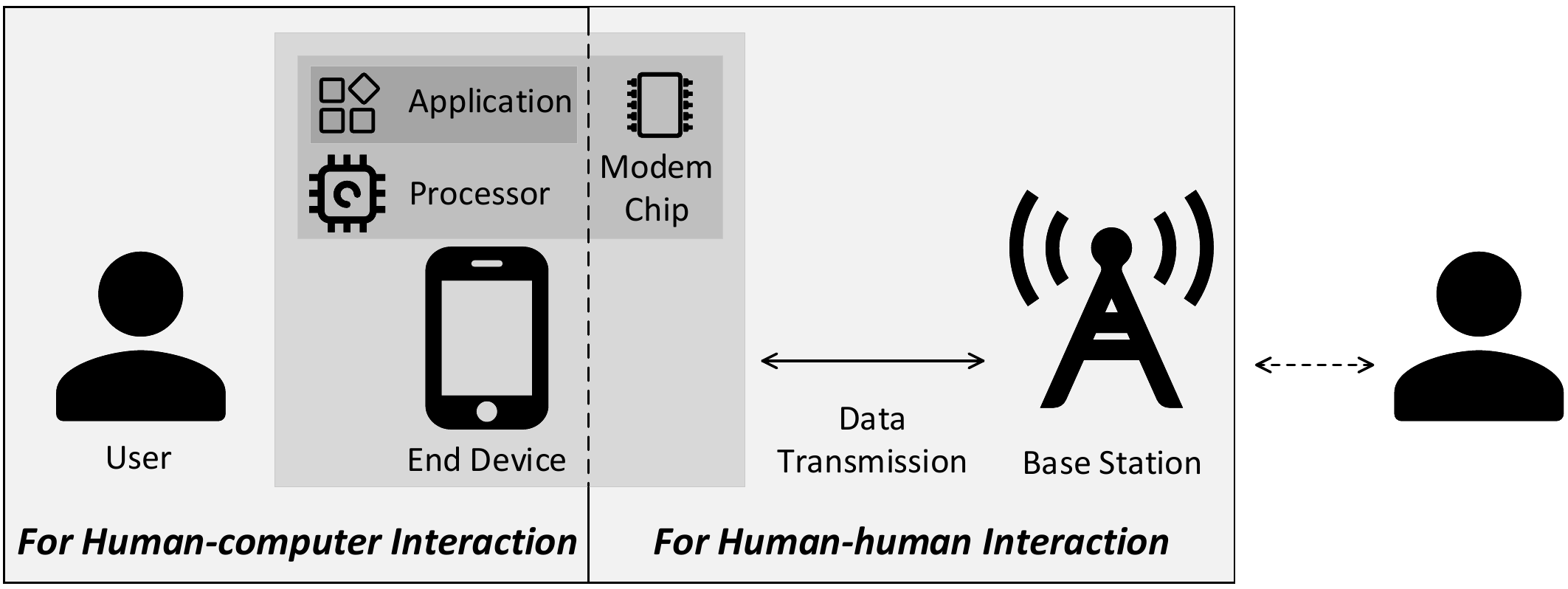}
	\caption{Key components for carbon reduction during the whole interaction process.}
	\label{fig:Interaction}
\end{figure}

Facing the severe carbon issue, it is urgent to develop green techniques to reduce carbon emissions from the interaction. As listed in Table~\ref{tab:InteractionGreenTechniques}, there have been many studies to improve energy efficiency of end-devices and applications for human-computer interaction, and networking equipment for human-human interaction. Figure~\ref{fig:Interaction} depicts these key components for carbon reduction during the whole interaction process. Similar to the PUE used to indicate the energy efficiency of datacenters, the indicator performance per watt (PPW) can be used to indicate the energy efficiency of a particular end-device or its internal hardware component. 

\subsection{Energy Efficiency Improvement of End-devices for Human-Computer Interaction}

An end-device generally contains various components like processors, memories, modem chips, etc., which dominate its total energy consumption~\cite{carroll2010analysis}. In addition to green techniques for saving energy directly, the heat harvesting technology, similar to the one used in datacenters, make sense to metaverse users for achieving a longer standby time of their end-devices. In the following, we will discuss green techniques on processors and the heat harvesting technology, and leave the discussion on modem chips, a key networking component in Section~\ref{sec:network}.

\subsubsection{Processors}\label{sec:hardwaredeviceprocessors}
Processors are core components for providing computing power in an end-device, including general-purpose processors (e.g., CPUs) and newly emerging special-purpose processors (e.g., GPUs, NPUs, DSPs). The latter are designed to execute only specific workloads, and is believed to have a higher energy efficiency.

\textbf{Heterogeneous processing architectures.} There are many techniques developed for general-purpose processors, such as the heterogeneous processing architecture technology and die shrink technology. As a representative architecture, the ARM big.LITTLE is widely used in modern mobile phones~\cite{ArmCore}. By equipping devices with both energy-efficient and high-performance heterogeneous processors, dynamic task allocation can be realized with this technology to maintain high performance while improving PPW. For the various kinds of metaverse workloads, developers should carefully decide on the best processor candidates. 

\textbf{Die shrink.} Although Moore's law may come to an end by around 2025, it is still effective to improve PPW by shrinking the die further~\cite{DieShrink}. Recently, Qualcomm has announced its Snapdragon 8 Gen 1 processor with the up-to-date 4~nm technology~\cite{SnapdragonChip}. Compared to its previous generation Snapdragon 888 with the older 7~nm technology, this new processor achieves 30\% power savings. Furthermore, it is reported that IBM has already created the world's first 2~nm processor in the lab, which can reduce as high as 75\% energy at the same performance as compared to the mainstream 7~nm processors~\cite{FirstChip}. When developing the end-devices for the metaverse, processors with smaller transistors can be preferentially selected to achieve not only higher energy efficiency but also better user experience although there may be potentially higher costs. For example, the 75\% energy savings achieved by the 2~nm technology also mean that the battery life can quadruple after the battery gets fully charged. This is especially significant for end-devices with limited battery capacity since users may not want to charge them halfway while enjoying the metaverse.

\textbf{Deploying special-purpose processors.} In addition to the above general-purpose processors, special-purpose processors have gained increasing attention from both academia and industry. These special processors have been recognized to improve energy efficiency significantly and even become the future of computing~\cite{FutureofComputing}. Considering the great benefits of the AI technique, AI accelerators are one of the most successful special-purpose processors. Recent works have designed various AI accelerators for enabling highly efficient AI inference on end-devices, such as Google's Edge TPU~\cite{EdgeTPU} and Samsung's Exynos 2100 processor~\cite{Exynos2100}. The later incorporates tri-cluster CPUs, a GPU, an NPU, etc., and the integrated NPU improves the energy efficiency to 0.84 mJ per inference~\cite{Exynos2100} from a typical value of tens to hundreds of millijoules~\cite{AndroidSoC}. Besides AI accelerators, other studies aim to design special-purpose processors tailored to metaverse devices. For example, Microsoft equips its AR headsets HoloLens~\cite{MicrosoftHoloLens} with a holographic processing unit that is responsible to process and integrate all the data streaming collected from on-board sensors. This custom processor consumes only less than 10~W, showing its high energy efficiency concerning the huge amount of workloads to be processed~\cite{HoloLens}. In all, as compared to general purpose processors, these special-purpose processors hold great promise for the metaverse since the end-devices mainly perform specific types of workloads, such as data collecting and AI inference.

\textbf{\underline{Implications}}~\textit{Similar to datacenters, it is necessary to deploy special-purpose processors on end-devices for processing various metaverse workloads efficiently. However, it is nontrivial for universal end-devices like mobile phones to balance performance, power demand, energy efficiency, die size, cost, and others when integrating these processors. Thus, more dedicated devices like VR headsets with VR chips need to be studied and developed beyond existing mobile phones.}

\subsubsection{Others}\label{sec:hardwaredeviceother}
\textbf{Heat harvesting.} Some external components are well studied to power end-devices through heat harvesting. Dai et al.~\cite{dai2018exploiting} present a TEG-based heat harvesting and reusing framework for mobile phones, and the generated electricity can help extend the battery life. Similarly, Yap et al.~\cite{yap2016thermoelectric} leverage TEGs for heat recovery from human bodies. Without any active heat input to TEGs, Raman et al.~\cite{raman2019generating} further show the potential to recycle heat in the dark. In summary, although the generated power is only a few to tens of mW at present, it is still meaningful to make further attempts on the heat harvesting technology, concerning the surprisingly high energy consumption of end-devices (presented in Section~\ref{sec:InteractionCarbon}) on the way to the metaverse.

\subsection{Energy Efficiency Improvement of Applications for Human-Computer Interaction}\label{sec:application}

On top of end-devices, there are various metaverse applications, such as virtual conferencing, virtual manufacturing, and cloud gaming. These applications rely on processing technologies such as AI inference and video analytics to generate the content, as well as display technologies including VR, AR, and MR to display the content. In the following, we will discuss green techniques on processing and display technologies.

\subsubsection{Processing technologies --- AI inference}\label{sec:application:processing}
As discussed in Section~\ref{sec:framework:carbon}, the AI technology plays an important role in building a virtual world and virtual persons in the metaverse and afterwards, operating the virtual world in a real-time manner. In addition to datacenter infrastructure, Facebook proposes to deploy AI inference on end-devices to achieve reduced latency and become independent of changeable network conditions~\cite{wu2019machine}, which makes sense to metaverse workloads with high requirements on QoE or privacy. Considering the limited computing power and battery capacity of end-devices, many green techniques have been advanced for AI inference. 
In the following, we will discuss these techniques from the perspective of the application and system levels. 

\textbf{AI model compression.} As a common application-level method, AI model compression has been evaluated by many works, and can be divided into several methods, including quantization, sparsification, tensor decomposition, and so on~\cite{deng2020model}. Qualcomm prompts the quantization method, where the addition and multiplication operations can reduce energy consumption by 30$\times$ and 18.5$\times$, respectively, and the memory access energy decreases by up to 4$\times$ by leveraging the INT8 formats instead of conventional FP32 formats~\cite{Quantization}. Yang et al.~\cite{yang2017designing} develop a specification method that achieves up to 3.7$\times$ energy reduction with less than 1\% top-5 accuracy loss. Although model compression shows the huge potential in saving energy, there can be perceptual accuracy drop. Depending on the type of metaverse applications and end-devices, developers should make decisions on different model compression methods and their combinations to achieve optimal energy efficiency and accuracy trade-offs.

\textbf{Collaborative execution.} Another application-level method is the device-edge collaborative execution proposed by Kang et al.~\cite{Neurosurgeon}. 
Considering that the energy consumption of end-devices comes chiefly from the computation and data transmission, the authors point out that the data transmission energy can be reduced by a large margin through partial offloading, i.e., model layers before the partition point are processed locally, while the rest are transmitted to the cloud/edge platform and processed remotely. The results show that mobile energy consumption can be reduced by 59.5\% on average and 94.7\% at most. Instead of a single partition point, Eshratifar et al.~\cite{eshratifar2019jointdnn} further consider several partition points depending on the DNN model structure, and achieve up to 32$\times$ mobile energy savings. However, although the collaborative execution can save energy of end-devices, there exists extra energy consumption of the cloud/edge platform which is neglected by prior works.

\textbf{\underline{Implications}}~\textit{Different from other techniques, collaborative execution involves cloud and/or edge that are/is responsible to process the second part of inference. However, more works are still necessary to investigate this additional part of energy consumption as compared to completely local execution.}

\textbf{Dynamic voltage frequency scaling.} The system-level technique DVFS is especially important to end-devices with limited battery capacity, and is well studied for years. Tang et al.~\cite{tang2019impact} investigate the impact of GPU DVFS settings on energy consumption of AI inference tasks. By selecting a proper GPU frequency, they achieve as high as 26.4\% energy reduction. Many recent works start to focus on a hybrid solution that considers both model compression and DVFS, which helps achieve different trade-offs. To achieve trade-offs between energy efficiency and accuracy, Nabavinejad et al.~\cite{nabavinejad2019coordinated} develop an approach to select proper model precision (e.g., INT8 and FP32) and GPU frequency, and the overall energy consumption decreases by up to 28\%. Bateni et al.~\cite{bateni2020neuos} leverage the hybrid solution to achieve a trade-off among energy efficiency, accuracy, and latency, and save the energy by up to 68\%. The hybrid solution provides much more candidates for metaverse developers. However, in view of the various metaverse applications and heterogeneous end-devices, it is nontrivial to customize the solution for each application-device combination.

\textbf{Workload scheduling.} As a widely-used system-level technique in both datacenters and end-devices, workload scheduling has been well studied. For end-devices, workload scheduling usually combines with model selection/compression and DVFS. Wang et al.~\cite{wang2021asymo} combine workload scheduling with DVFS in view of the widely deployed asymmetric multiprocessors on end-devices, e.g., ARM-based multicore CPUs, and devise an asymmetry-aware partitioning and task scheduling solution. 
Wan et al.~\cite{wan2020alert} combine application-level model selection and system-level resource allocation. As compared to approaches at either the application or system level, this hybrid solution reduces the energy consumption by over 13\%. 

\textbf{Customized AI inference frameworks.} Another system-level method is developing customized AI inference frameworks, which is of great importance to AI inference-based metaverse applications. Based on TensorFlow, Google develops the TensorFlow Lite framework~\cite{TensorFlow} to enable efficient AI inference on end-devices with computing power and energy constraints. Zhang et al.~\cite{zhang2019openei} further devise a framework combined with application-level optimization, such as adaptive model selection. Since these methods improve energy efficiency usually at the expense of performance, metaverse developers should carefully select the components and its frequency that achieves the best energy efficiency and performance trade-offs. 

\subsubsection{Processing technologies --- video analytics}\label{sec:application:video}
Video analytics leverages the AI technique to complete various computer vision tasks for videos, such as motion detection and behavior tracking~\cite{VideoAnalytics}. Therefore, many green techniques can be developed based on the properties of video streaming. Feng et al.~\cite{feng2021palleon} establish a runtime system to select a compressed AI model dynamically based on the time locality and class skew properties of video streaming, which achieves up to 6.7× energy savings. Song et al.~\cite{song2020vr} point out that the video frames can be categorized into three types, i.e., I, P, and B frames. Since B frames can be completely reconstructed by I and P frames, AI inference-related computations for the B frames can be significantly reduced. These video-oriented techniques are key to the streaming-based metaverse applications from virtual conferencing to virtual manufacturing.

\subsubsection{Display technologies}\label{sec:application:VRAR}
The VR, AR, and MR technologies bridge the virtual and physical worlds, and provide immersive experience to users. Specifically, VR displays a completely virtual world through 360$^\circ$ video streaming. In comparison, AR needs to place virtual objects in the user's field of view properly after recognizing the surrounding physical environment~\cite{MixedReality, MixedReality2}. MR is regarded as a combination of VR and AR. 

\textbf{Reducing unnecessary data transferring and redundant calculations.} To satisfy the strict latency requirements of real-time interaction, current VR and AR devices mainly perform computational tasks locally. 
For VR devices that support 360$^\circ$ video streaming, there are abundant computations and data movements between GPUs, DRAMs, and display panels, which wastes huge amounts of energy. Haj-Yahya et al.~\cite{haj2021burstlink} focus on a new data transferring architecture that demonstrates up to 33\% energy savings. Zhao et al.~\cite{zhao2020deja} make full use of temporal and spatial correlations of the video to reduce unnecessary computations, which lowers the energy consumption by 17\%. 
For AR devices, object recognition is a key component for the environmental perception. Apicharttrisorn et al.~\cite{apicharttrisorn2019frugal} design an object tracking solution to improve energy efficiency while maintaining high detection accuracy by reducing redundant computations. Their real-world experiments show up to 73.3\% energy reduction. In view of the popularity of VR, AR, and MR in the metaverse and the increasing bit rates of video frames, there is likely to be huge and probably redundant data transmission, as well as redundant computations for AR applications to conduct AI inference in the background. It is urgent for metaverse developers to adopt more green techniques and put them into practice, while preventing perceptible QoE drops.


\subsection{Energy Efficiency Improvement of Networking for Human-Human Interaction}\label{sec:network}

Although 5G provides many attractive properties desirable to the metaverse, the high energy consumption for transmitting data among end-devices and base stations can generate large carbon emissions as presented in Section~\ref{sec:InteractionCarbon}. In the following, we will discuss green techniques on modem chips in end-devices, base stations, and data transmission. 

\subsubsection{Modem chips in end-devices}\label{sec:networkDevices}
The energy consumption of end-devices for data transmission comes chiefly from modem chips. There are several green techniques on modem chips at hardware and software levels. 

\textbf{Using integrated modems and dynamic voltage frequency scaling.} At the hardware level, with the evolving of the chip technology, instead of building separate 5G modems, MediaTek proposes its Dimensity series with integrated modems to reduce the energy consumption~\cite{5GMorePower}. The adopted UltraSave technology~\cite{5GModems} further saves energy by adjusting the chip frequency dynamically. 

\textbf{Power management.} At the software level, to learn the energy share of each networking-related operation, it is necessary to build power models in the first place and then enable power management. In particular, according to a power model developed by 3GPP~\cite{5GDeviceEfficiency}, the control channel monitoring operation dominates the total energy consumption (e.g., over 50\%) although this operation only occupies a small portion of time cycles (e.g., about 5\%) without any data transmission. This monitoring operations could be reduced on metaverse devices in view of the continuous and stable connection to the virtual world in VR/AR scenarios. For end-devices using different frequency bands, such as mmWave and sub-6 GHz, however, the energy shares of each operation behave differently~\cite{5GDeviceEfficiency}. This indicates that it would be useful to design custom-designed power management schemes for them, such as discontinuous reception configuration settings recommended by 3GPP~\cite{3GPP5G}, and dynamic search space adaptation proposed by Ericsson~\cite{5GDeviceEfficiency}. 

\textbf{Switching between wireless technologies.} It is recognized that it is not always efficient to use the 5G signal for communication rather than 4G especially when the throughput is low~\cite{xu2020understanding, narayanan2021variegated}. Xu et al.~\cite{xu2020understanding} advance a dynamic mode switching solution that switches the device to 5G only when the throughput demand is beyond 4G's capacity. The simulation results show that the energy consumption can be reduced by up to 24.8\% as compared to using 5G all the time. Narayanan et al.~\cite{narayanan2021variegated} propose a solution for video streaming applications which switches the device to 4G when the predicted 5G throughput is below a threshold. However, switching between 4G and 5G frequently can incur hiccups, which may greatly degrade user experience when users are immersing themselves in virtual roles. Besides, the switches themselves consume significant amounts of energy, which would decrease the benefits from 4G-5G switches. Recently, Kim et al. develop a new material of switches for 5G devices~\cite{kim2020analogue} and even 6G ones~\cite{kim2022monolayer}. The ones for 5G not only deliver a 50$\times$ increase in energy efficiency but also realize as high as nanosecond switching speeds, which is promising for the metaverse with continuous multiplayer interaction. 

\subsubsection{Base stations}\label{sec:networkStations}
There are studies focusing on architectural design~\cite{chih2020energy} and power management~\cite{Huawei5GPower, wu2015energy, suzuki2021dynamic} to save energy of base stations. For metaverse applications, the power management can be one of the key considerations when transferring data between users. Recently, China Tower and Huawei propose their 5G Power solution that enables intelligent peak shaving, intelligent voltage boosting, and intelligent energy storage~\cite{Huawei5GPower} in an end-to-end manner, which helps save significant costs during retrofitting existing low-capacity site infrastructure. Evaluation results show that this solution is able to save around 4,130~kWh of electricity, and 1,125~kg of carbon emissions per site per year. 

\subsubsection{Data transmission}\label{sec:networkTransmission}
During data transmission, similar to heat harvesting for end-devices, it also makes sense to recycle energy from the high-frequency radio signals of 5G~\cite{imran2019energy}. Wu et al.~\cite{wu2017overview} further summarize some schemes to improve energy harvesting, such as energy beamforming and channel state information acquisition. 

\textbf{\underline{Implications}}~\textit{Different from conventional content delivery network (CDN) that distributes just texts, images, and HD or UHD videos, many metaverse applications are built upon extremely high-bitrate videos, such as 16K panoramic video-on-demand and volumetric video streaming~\cite{zhang2021deepvista, han2020vivo}, which will inevitably bring much higher energy consumption and carbon emissions. Thus, it is urgent to conduct more quantitative measurements on these newly emerging applications under real environment and develop targeted green techniques on the way to the metaverse.}


\subsection{Directions to Become Green}\label{sec:interactionChallenge}
Although there have been extensive studies on green techniques, there are still several challenges in the interaction layer. Specifically, we propose two insights into the selection of green techniques and heavy network overhead of the existing CDN technology. 

(1)~\textit{\textbf{Insight:}} \textbf{It is nontrivial to select the best green technique and settings for each metaverse end-device and application.} 
As a large number of new types of metaverse devices and applications start to emerge, such as smart gloves, existing green techniques lack generalizability and scalability. For example, the effectiveness of the AI model compression technique depends on the hardware architecture~\cite{deng2020model}. That is, an AI model compressed for one hardware type is likely to become less efficient and even inefficient for another, especially the newly-emerged ones with different architectures. However, this will increase the burden of developers as they need to have a rich understanding of various devices and applications and manually select the best green techniques and parameters for each device and application. 

\textit{\textbf{Direction:}} To reduce the burden of metaverse developers, it will be promising to develop a set of automated search engine tools based on intelligent algorithms like reinforcement learning. With such a tool, developers only need to provide necessary inputs like the workload type, hardware type and performance requirements, and the tool can automatically provide suggestions and recommend the best green techniques and parameters.

(2)~\textit{\textbf{Insight:}} \textbf{The network overhead of the CDN technology gets worse as metaverse users are distributed all over the physical world.} It is recognized that CDN can help distribute contents (e.g., 3D background images) based on user preferences in different locations. However, under the scope of the metaverse, users are usually scattered across the physical world, even if they are very close in the virtual world. To this extend, CDN nodes need to cache contents for each user far from each other, which greatly increases the network transmission burden. 

\textit{\textbf{Direction:}} Since metaverse users are typically far from each other in the physical world, green techniques based on edge computing are promising to reduce the network overhead and potential carbon emissions. For example, we can transmit only low-quality frames from the cloud and then perform super-resolution at the edge. In addition, instead of downloading all rendered frames from the cloud, edge nodes can generate subsequent frames based on historical frames and their temporal correlation. The above solutions trade the computation for communication, so developers should ensure a low computing overhead. 

\section{How can the Economy Become Green?}\label{sec:payment}

\begin{table*}[t]
	\renewcommand{\arraystretch}{1.5}
    \centering
    \caption{Green Techniques for the Economy.}
    \begin{tabular}{|c|l|l|}
    \hline
    \textbf{Level}              & \multicolumn{1}{c|}{\textbf{Green Technique}}      & \multicolumn{1}{c|}{\textbf{Work}}       \\ \hline
    \multirow{2}{*}{\makecell*[c]{Blockchain (Sec.~\ref{sec:paymentSolution})}} & Consensus protocol design and selection               & Proof of Solution~\cite{chen2022blockchain}, Bada et al.~\cite{bada2021towards}                \\ \cline{2-3} 
            & Transaction combination                    & Lightning Network~\cite{lightning-network}                  \\ \hline
    \end{tabular}
    \label{tab:PaymentGreenTechniques}
\end{table*}

Similar to the physical world, the metaverse economy requires a financial system to ensure trusted transactions between users and verify their ownership rights. In view of the requirement of high security, the financial system of the metaverse has widely incorporated the blockchain technology. As a virtual currency, the cryptocurrency is a necessary medium for online transactions, and non-fungible tokens (NFTs) are used to prove the ownership rights of users' virtual properties~\cite{NFTsmetaverse}. For example, many metaverse platforms, such as Decentraland, Somnium Space, Nextech AR, etc., have allowed users to buy and sell their virtual properties with blockchain-based cryptocurrencies~\cite{nft-decentraland, nfthouse, NextechNFT}. 

The blockchain technology, dating back to 2008, refers to an immutable distributed ledger on a peer-to-peer network which ensures trusted transactions or records with a consensus protocol~\cite{bitcoinwhitepaper, herlihy2019blockchains}, and achieves adequate security through its natural features as follows. The blockchain consists of a chain, i.e., a sequence of blocks, with the last block containing the hash value of its former block. Thus, a minor modification to a block will change its hash value and thus break the chain. As each worker in the peer-to-peer network keeps a copy of the chain that is synchronized frequently, no one can tamper with data on the chain without paying a huge price (e.g., owning more than half of the total computing power). In despite of the security property of the blockchain, it can cause significant carbon emissions for computing. In the following, we will introduce the carbon issue of the economy layer with the blockchain technology, and then analyze green techniques to improve its energy efficiency. In the end, we provide future directions to become green.

\subsection{The Carbon Issue of the Economy}\label{sec:paymentBlockchainEnergy}
Although the blockchain technology is desirable to the metaverse, there exists a severe carbon issue. For example, Bitcoin is known for its large amount of carbon footprints on useless hash calculations. According to a study~\cite{CryptocurrencyCambridge}, Bitcoin is estimated to consume 126 TWh of electricity in 2021. Since Bitcoin occupies about 68.39\% of cryptocurrency mining energy~\cite{gallersdorfer2020energy}, the carbon emissions by all the blockchains for monetary transactions can be roughly estimated as 77 Mt in 2022 and 460 Mt in 2030~\cite{CryptocurrencyCambridge}. 
The main reason for the high carbon emissions is that Bitcoin uses proof of work (PoW) as its consensus protocol. All workers in the peer-to-peer network can create a new block, and each of them races to solve a complex cryptographic hash puzzle in order to add its own block to the blockchain and get a reward. This process leads to many useless hash computations and thus high carbon emissions. 
Besides, the problem still remains serious for Ethereum although it has started to replace PoW with more efficient proof of stake (PoS). For example, the carbon footprint of an Ethereum transaction is comparable to about 329,000 credit card transactions currently~\cite{EthereumvsVisa}.

\begin{figure}[t]
	\centering
	\includegraphics[width=1\columnwidth]{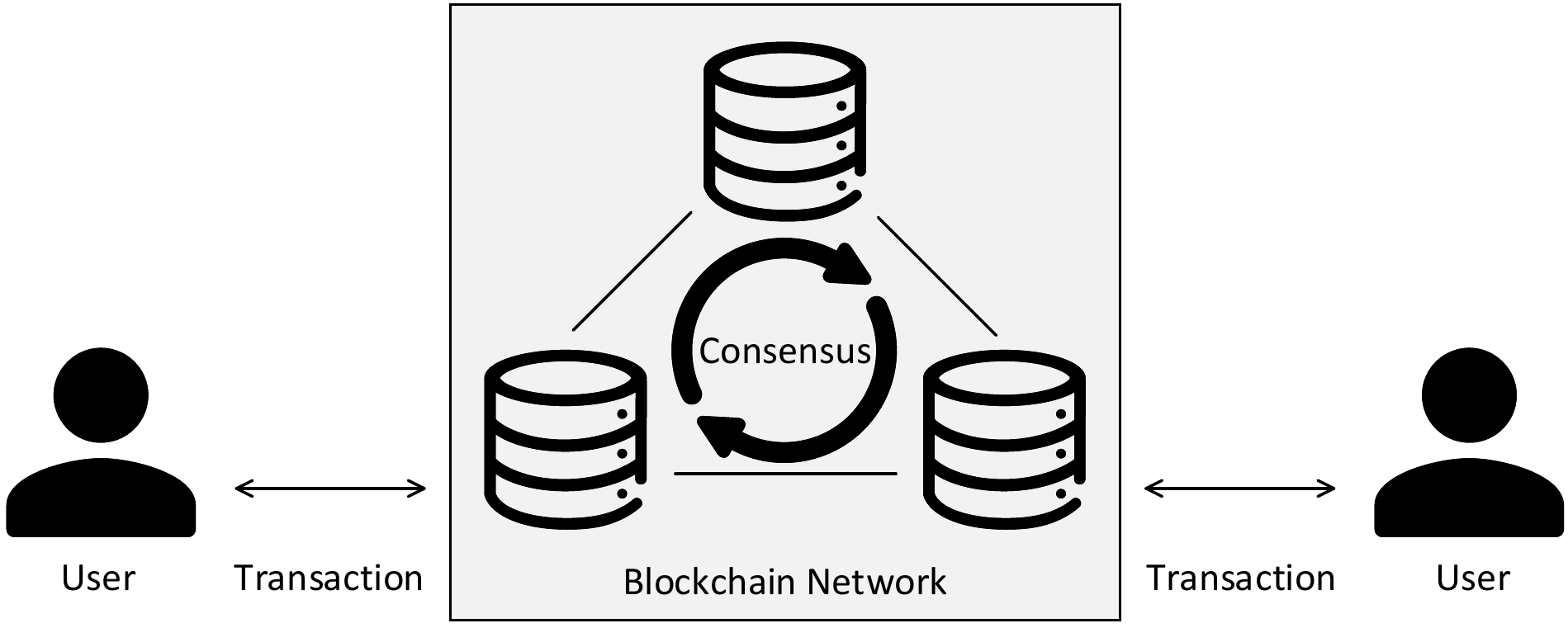}
	\caption{Key components for carbon reduction during the economy process.}
	\label{fig:Payment}
\end{figure}

Therefore, it is urgent to develop green techniques to reduce carbon emissions from the economy layer. As listed in Table~\ref{tab:PaymentGreenTechniques}, there are green techniques for improving energy efficiency and reducing carbon emissions of the blockchain. Figure~\ref{fig:Payment} shows the key components for carbon reduction of the economy process. 

\subsection{Energy Efficiency Improvement of the Blockchain}\label{sec:paymentSolution}
Many researchers have made efforts on more efficient consensus protocols in the system level and transaction management in the application level.

\subsubsection{Consensus protocol design and selection}\label{sec:paymentSolutionProtocol}
Consensus protocols play a critical role in ensuring the security~\cite{zhang2020analysis}, while higher security usually brings higher energy consumption. Many blockchain platforms, including Bitcoin, Ethereum, etc., adopt different consensus protocols, such as PoW and PoS, with different trade-offs between security levels and the amount of computations, and the latter further determines the energy consumption and carbon emissions. As presented in Section~\ref{sec:paymentBlockchainEnergy}, PoW has an enormous appetite for energy. To achieve an energy-efficient consensus protocol, Sunny King et al.~\cite{king2012ppcoin} design PoS in PeerCoin. Rather than taking lots of computing power to perform complicated hash calculations, PoS requires workers to place their coins at stake and chooses the block creator by an algorithm based on each worker's stake. Many blockchains and metaverse platforms have adopted PoS to improve efficiency although its security level is lower than PoW. For instance, Ethereum is gradually migrating from PoW to PoS, which will cut the energy consumption by more than 99\%~\cite{EthereumBlog}. Some metaverse platforms like Decentraland and Somnium Space have chosen to build their financial systems based on Ethereum. 

Recently, various consensus protocols have been proposed with different security levels and energy demands. Daniel proposes delegated proof of stake (DPoS), a variant of PoS~\cite{larimer2014delegated}. It allows workers who hold a stake to vote to elect the block creator, which is proved to achieve higher energy efficiency~\cite{zhang2020analysis}. In particular, Eleonora et al.~\cite{Eleonora2020blockchain} estimate the energy consumption of PoW, PoS, and DPoS. The results show that the annual energy consumption of DPoS is the lowest (about 1.2~GWh per year), PoS is sightly higher than DPoS (about 7.3~GWh per year), and both are significantly lower than PoW (about 33~TWh per year).  Chen et al.~\cite{chen2022blockchain} propose Proof of Solution (PoSo), a new blockchain consensus protocol where the workers reach a consensus by solving a meaningful optimization problem rather than useless hash calculations as PoW. This could be very useful in the metaverse as there are many optimization problems whose solutions are hard to calculate but easy to verify, such as recognizing the shortest path for guiding the users in the virtual world. 

Provided with so many consensus protocols, it is not easy to select the best one for a specific scenario. Bada et al.~\cite{bada2021towards} critically investigate 18 consensus protocols with different estimated energy consumption and propose a framework for selecting a proper consensus protocol. This can be a useful tool for metaverse service providers to realize flexible trade-offs between security levels and energy consumption.

\subsubsection{Transaction combination}\label{sec:paymentSolutionLightning}
Combining multiple transactions together can decrease the number of transactions recorded in the blockchain and thus reduce the carbon emissions. Poon et al.~\cite{lightning-network} describe Lightning Network in a white paper for improving Bitcoin's scalability. Lightning Network allows users to make multiple transactions together and records them as a single transaction by creating a micropayment channel between users. Recently, the Ethereum platform has created its own Lightning Network named Raiden Network~\cite{Raiden-Network}. This technique would be desirable to the metaverse as both the demand for transactions and the transaction volume continue growing.

\subsection{Directions to Become Green}\label{sec:paymentChallenge}
Although there have been extensive studies on green techniques, there are still several challenges in the economy layer. Specifically, we propose two insights into handling frequent yet small transactions, and the trade-off between security and carbon emissions. 

(1)~\textit{\textbf{Insight:}} \textbf{It is rather inefficient to conduct a huge amount of transactions frequently for the metaverse.} 
Unlike the physics world, the transaction process in the metaverse relies extensively on the blockchain technology. However, the high-volume, frequent yet small transactions pose a serious problem on efficiency if all the transaction processes are recorded on the blockchain. For example, a Bitcoin transaction consumes up to 1,173 kWh~\cite{CryptoEnergy}.

\textit{\textbf{Direction:}} In fact, it is not always necessary to record all transactions on the blockchain in real time. Instead, we can merge consecutive and relative transactions in advance before adding them to the chain while ensuring the correctness.

(2)~\textit{\textbf{Insight:}} \textbf{It is nontrivial to achieve a right trade-off between security and carbon emissions.} It is no secret that security is the most important consideration for the economic system. New consensus protocols with fewer carbon emissions often partially sacrifice the security. What is worse, to save costs, some platforms even choose to store users' property just on the platform rather than the blockchain, which reduces carbon emissions by adding only a few blocks on the chain but poses security threats. Recently, an NFT owned by a famous star Jay Chou has been stolen, probably due to the malicious attack to the NFT platform without enough security guarantees~\cite{NFTstolen}.  

\textit{\textbf{Direction:}} To achieve a right trade-off between the security and carbon emissions, it is important to enforce different security levels based on the transaction types (e.g., proof of ownership, large-value transactions, daily small-value transactions) by adopting different consensus protocols. For example, large-value transactions have higher security requirements, but the transaction frequency is low, in which case PoW could be more suitable. On the contrary, for daily small-value transactions with a high transaction frequency, PoS could be better as the requirement for security is lower. However, the mashup of various consensus protocols place a higher requirement on the management of the blockchain system at the same time.




\section{A Low-Carbon Metaverse from the Governance Perspective}\label{sec:economic}

\subsection{Public Policies in the Physical World}\label{sec:policy}

In this section, we briefly discuss several public policies to spur energy efficiency and carbon neutrality investments, such as green financing and investing, carbon-neutral bond issuance, carbon trading, and carbon pricing. At the beginning of 2021, China issued its first batch of carbon-neutral bonds that would support carbon-reduction projects~\cite{CarbonBonds}. Similarly, as the first greenhouse gas emissions trading system founded in 2005, the European Union Emissions Trading System (EU ETS) aims at greenhouse gas emission reduction under the ``cap and trade" principle~\cite{EUETS}. Companies regulated by the EU ETS must restrict their carbon emissions within the carbon allowances, otherwise they need to buy these allowances from other companies on the carbon market. The EU ETS has already been proved to be effective and achieves up to 3.8\% of EU-wide emission reduction between 2008 and 2016 based on an estimation compared to carbon emissions without the EU ETS~\cite{bayer2020european}. Besides Europe, many other countries and regions have built their carbon emission trading systems~\cite{ICAP}, such as the US, the UK, China, etc. However, no specific policies have been developed for the metaverse ecosystem currently, which should be put on the agenda. 


\subsection{Regulation of Users in the Virtual World}\label{sec:regulation}
Although users begin to live in a virtual world and everything gets to be unreal, all their actions will generate a certain amount of carbon emissions in the physical world as well. However, for a user activity in the virtual world, its carbon footprint will vary greatly from that occurs in the physical world. For example, doing exercise in the virtual world can lead to relatively high carbon emissions for rendering consecutive video frames, but it only consumes our own body's energy in the physical world. Another example is the way of travelling. In the physical world, hiking only consumes the user's own energy, while flying will bring a lot of carbon emissions. By comparison, due to the need to render a sequence of video frames for a long time, hiking in the virtual world can bring high carbon emissions. Instead, a low-carbon way is to ``take a virtual plane'', that is to say that the user can directly click a button to arrive at his destination. In this way, travelling all over the world incurs negligible carbon emissions as no real vehicles are needed and only the background image needs to be replaced on the screen.

In view of the difference, \textit{a new assessing, recording, and regulating mechanism} tailored to the metaverse is necessary. First, we need a new strategy to assess carbon emissions of users' activities and record them through carbon management agencies in the virtual world. Then, we can recommend activities to users based on the recorded carbon information of each activity. Finally, to stimulate users to be aware of their carbon footprints in the metaverse, we can charge them in the physical world or just provide them with a virtual bonus in the virtual world. We further propose a quantitative indicator \textbf{Carbon Utility (CU)}, defined as the ratio of service quality brought by activities in the virtual world to the produced carbon emissions in the physical world. This value is affected by the type of activities (e.g., travelling and holding a party), the implementation (e.g., 4K video, 360$^\circ$ video, and real-time multiplayer interaction), and the energy efficiency of each component in each layer (e.g., datacenters and end-devices). This indicator not only provides information on the carbon bottleneck (i.e., the components with the largest carbon emissions), but also helps the metaverse platform recommend ``green'' activities to users in the virtual world, and ``green'' implementations to developers in the physical world.

\textbf{\underline{Implications}}~\textit{As the carbon footprint of the same user activity behaves totally differently between in the physical world and virtual world, it is necessary to develop a new assessing, recording, and regulating mechanism on the carbon impacts of these online user activities.}



\section{Conclusion and Future Directions}\label{sec:conclusion}
With the fast evolvement of the metaverse, it is believed that a growing number of users will join the metaverse when it provides better QoE of working, playing, trading, etc., in the near future. However, we argue that this probably comes at the expense of huge energy consumption and carbon emissions, which will largely hinder the way to carbon neutrality. To better understand this carbon issue, we first split the metaverse into three carbon-intensive layers and estimate their carbon footprints from 2022 to 2030. Our results show that the carbon emissions of the metaverse in 2030 will reach nearly 0.5\% of the global carbon emissions if we cannot take effective measures. In view of this critical issue, we then present a wide range of current and emerging green techniques for reducing carbon emissions and analyze their limitations when facing the specific requirements of metaverse workloads. Finally, we propose several insights and future directions to help make each of the layers become green.

As discussed in Section~\ref{sec:framework}, the metaverse is built upon a variety of existing technologies. For the metaverse as a whole, each existing green technique still focuses on a single or a few components within a layer, and only takes into account limited performance metrics. We argue that this is no more efficient since the metaverse is an all-inclusive world involving extensive components. For example, the device-edge collaborative execution solution for AI inference is demonstrated to be effective in reducing the device energy but only focuses on end-devices themselves. However, extra energy will be consumed by datacenter servers, switches, etc., which can be large but is neglected by existing literature. Another example is the workload consolidation, which is a general-use approach in datacenters to save the energy of IT systems by reducing the number of active servers. However, it can incur significant hotspots and thus increase the cooling energy consumption. 
To tackle the issue, we summarize three future directions to improve the energy efficiency of the metaverse as a whole in the following. 

(1) \textit{Performance metrics to be jointly considered:} as a higher energy efficiency can lead to various performance degradation of the metaverse, such as a lower computing speed and a lower security level, we need to consider multiple necessary performance metrics like latency, security, and cost at the same time to achieve a desirable trade-off. However, it is nontrivial to reduce the prohibitively large search space to get an appropriate solution. 

(2) \textit{Layers and components to be jointly considered:} instead of each individual component within a layer on which most ongoing efforts focus, such as processors or cooling equipment, it is necessary to focus on an end-to-end solution that jointly considers all the components and even all the layers to maximize the overall efficiency. To this extend, the first step is to capture the carbon footprint of each component in each layer from a specific metaverse application; the second step is to evaluate the energy efficiency of these components based on the indicators like PUE and PPW; the last step is to feed back, i.e., determine the best green techniques for the application according to the connections between these components, their priorities, the QoE requirement, etc. Since this requires a full knowledge of all the components in the metaverse, training metaverse engineers and developers is the key to the future.


(3) \textit{A new regulation mechanism on carbon footprints by the virtual world:} since the carbon footprints of various user activities in the virtual world behave differently from that in the physical world, we need to design a new assessing, recording, and regulating mechanism on carbon footprints. Particularly, we propose a quantitative indicator for the metaverse --- Carbon Utility, to reflect the carbon intensity of different user activities in the metaverse. Leveraging this indicator, future efforts are required on how to evaluate the service quality and carbon impacts of each user activity.

\ifCLASSOPTIONcaptionsoff
  \newpage
\fi

\def\UrlBreaks{\do\/\do-}
\bibliographystyle{IEEEtran}
\bibliography{refs}

\begin{IEEEbiography}[{\includegraphics[width=1in,height=1.25in,clip,keepaspectratio]{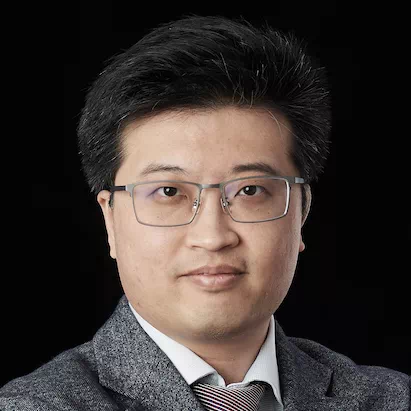}}]{Fangming Liu}
(S'08, M'11, SM'16) received the B.Eng. degree from the Tsinghua University, Beijing, and the Ph.D. degree from the Hong Kong University of Science and Technology, Hong Kong. He is currently a Full Professor with the Huazhong University of Science and Technology, Wuhan, China. His research interests include cloud computing and edge computing, datacenter and green computing, SDN/NFV/5G and applied ML/AI. He received the National Natural Science Fund (NSFC) for Excellent Young Scholars, and the National Program Special Support for Top-Notch Young Professionals. He is a recipient of the Best Paper Award of IEEE/ACM IWQoS 2019, ACM e-Energy 2018 and IEEE GLOBECOM 2011, the First Class Prize of Natural Science of Ministry of Education in China, as well as the Second Class Prize of National Natural Science Award in China.
\end{IEEEbiography}

\begin{IEEEbiography}[{\includegraphics[width=1in,height=1.25in,clip,keepaspectratio]{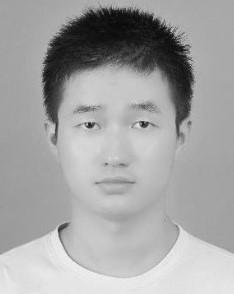}}]{Qiangyu Pei}
received the BS degree in physics from the Huazhong University of Science and Technology, China, in 2019. He is currently working toward the PhD degree in the School of Computer Science and Technology, Huazhong University of Science and Technology. His research interests include edge computing, green computing, and deep learning.
\end{IEEEbiography}


\begin{IEEEbiography}[{\includegraphics[width=1in,height=1.25in,clip,keepaspectratio]{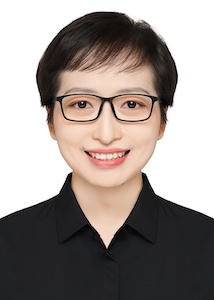}}]{Shutong Chen}
received her B.Sc. degree in the College of Mathematics and Econometrics, Hunan University, China. She is currently a Ph.D. student in the School of Computer Science and Technology, Huazhong University of Science and Technology, China. Her research interests include edge computing, green computing, and datacenter energy management.
\end{IEEEbiography}

\begin{IEEEbiography}[{\includegraphics[width=1in,height=1.25in,clip,keepaspectratio]{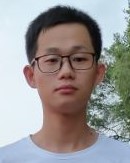}}]{Yongjie Yuan}
received the B.Eng. degree from the School of Computer Science and Technology, Huazhong University of Science and Technology, China, in 2021, where he is currently pursuing the M.Eng. degree. His research interests include edge computing and serverless computing.
\end{IEEEbiography}

\begin{IEEEbiography}[{\includegraphics[width=1in,height=1.25in,clip,keepaspectratio]{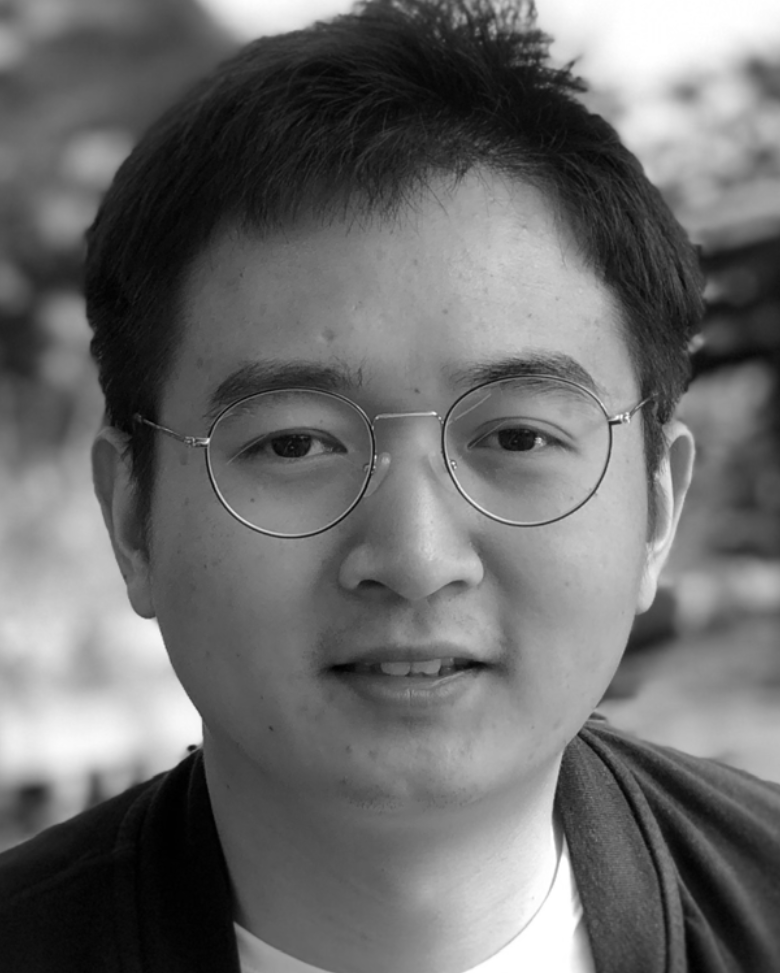}}]{Lin Wang}
received the Ph.D. degree from the Institute of Computing Technology, Chinese Academy of Sciences, in 2015. He held positions at Technical University of Darmstadt, SnT Luxembourg, and IMDEA Networks Institute. He is currently an Assistant Professor with the Computer Systems Section, Vrije Universiteit Amsterdam. His research interests include distributed systems and networking. He received an Athene Young Investigator Award from Technical University of Darmstadt in 2018 and a Google Research Scholar Award in 2022.
\end{IEEEbiography}

\begin{IEEEbiography}[{\includegraphics[width=1in,height=1.25in,clip,keepaspectratio]{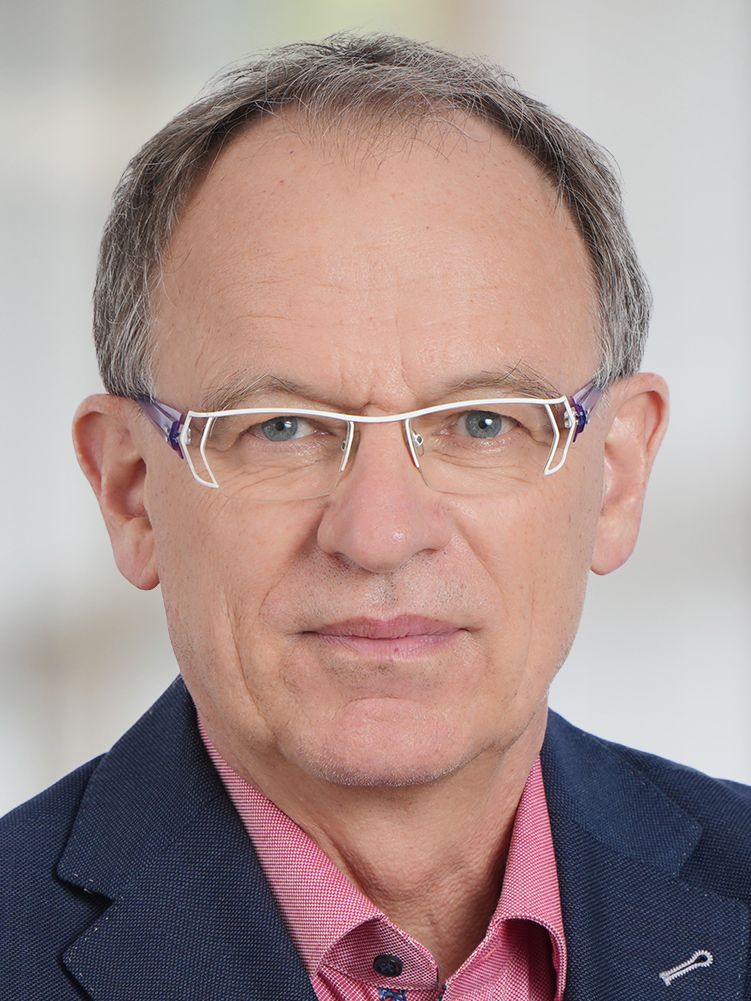}}]{Max M{\"u}hlh{\"a}user} is currently a Full Professor with Technical University of Darmstadt, where he is also the Head of the Telecooperation Laboratory. He holds key positions in several large collaborative research centers and is leading the Doctoral School on Privacy and Trust for Mobile Users. He and his lab members conduct research on the future Internet, human–computer interaction and cybersecurity, and privacy and trust. He founded and managed industrial research centers and worked as either a Professor or a Visiting Professor at universities in Germany, USA, Canada, Australia, France, and Austria. He is a member of acatech, the German Academy of the Technical Sciences. He was and is active in numerous conference program committees, as an Organizer for several annual conferences and as a member of editorial boards or a Guest Editor for journals, such as IMWUT, Pervasive Computing, ACM Multimedia, and Pervasive and Mobile Computing. He is a Senior Member of IEEE and a Distinguished Member of ACM.
\end{IEEEbiography}




\end{document}